\documentclass[showpacs,amsmath,amssymb,aps,pre,twocolumn,]{revtex4-1}
\usepackage{graphicx}
\usepackage{dcolumn}
\usepackage{bm}
\usepackage[breaklinks,colorlinks = true,linkcolor = red,urlcolor=blue,citecolor=red]{hyperref}
\usepackage{multirow}
\usepackage{array}
\usepackage{booktabs}
\usepackage{ctable}
\usepackage{ulem}
\usepackage{upgreek}
\usepackage{epsfig}
\usepackage{mathrsfs}
\usepackage{amssymb}
\usepackage{amsbsy}
\usepackage{color}
\usepackage{cancel}
\usepackage{pifont}
\usepackage{marginnote}
\usepackage{float}
\usepackage{verbatim}
\usepackage{subfigure}
\usepackage{bbm, dsfont}
\usepackage{lineno}
\usepackage{amsmath}
\definecolor{dgreen}{rgb}{0,0.7,0}

\makeatletter

\newcommand{\Rmnum}[1]{\expandafter\@slowromancap\romannumeral #1@}
\makeatother

\usepackage[utf8]{inputenc}

\begin{abstract}
We investigate the local time $(T_{loc})$ statistics for a run and tumble particle in an one dimensional inhomogeneous medium. The inhomogeneity is introduced by considering the position dependent rate of the form $R(x) = \gamma \frac{|x|^{\alpha}}{l^{\alpha}}$ with $\alpha \geq 0$. For $\alpha =0$, we derive the probability distribution of $T_{loc}$ exactly which is expressed as a series of $\delta$-functions in which the coefficients can be interpreted as the probability of multiple revisits of the particle to the origin starting from the origin. For general $\alpha$, we show that the typical fluctuations of  $T_{loc}$ scale with time as $T_{loc} \sim t^{\frac{1+\alpha}{2+\alpha}}$ for large $t$ and their probability distribution possesses a scaling behaviour described by a scaling function which we have computed analytically.  In the second part, we study the statistics of $T_{loc}$ till the RTP makes a first passage to $x=M~(>0)$. In this case also, we show that the probability distribution can be expressed as a series sum of  $\delta$-functions for all values of $\alpha~(\geq 0)$ with coefficients appearing from appropriate exit problems. All our analytical findings are supported with the numerical simulations.
\end{abstract}

\begin{document}

\title{Local time for run and tumble particle}
\author{Prashant Singh and Anupam Kundu}
\email{prashant.singh@icts.res.in}
\email{anupam.kundu@icts.res.in}
\affiliation{International Centre for Theoretical Sciences, Tata Institute of Fundamental Research, Bengaluru 560089, India}

\date{\today}

\maketitle
\section{INTRODUCTION}
Local time refers to the total time that a particle spends in the neighbourhood of a given point in space when it is evolved till time $t$. The study of the properties of local time comes handy in a wide range of interdisciplinary settings. For example, in chemical reactions where a catalytic agent reacts with reactants that are heterogeneously distributed in the space, the yield of the product is correlated to the amount of time spent by the agent in the vicinity of the reactants \cite{Wilemski1973,Benichou2005,Doi1975,Temkin1984}. Similarly, in biological applications, the action of a molecule inside a cell depends on the time that it spends inside the cell. This time is called the residence time which in the limit of small cell volume reduces to the local time \cite{Redner2001,Pal2019} . Therefore, it is important to compute the statistical properties of the local time for various processes. For stochastic processes, deciphering the local time statistics turns out to be important as it can provide the information about the spatio-temporal properties of the particle's trajectory. In the past, the local time has been widely studied in the context of various stochastic processes like diffusion in random potential landscape \cite{MajumdarSNC2002,SabhapanditS2006}, Ornstein-Uhlenbeck process \cite{Kishore2020}, continuous time random walk \cite{Carmi2010}, Brownian excursions \cite{Louchard1984}, uniform empirical process \cite{Csorgo1999}, reflected Brownian motion \cite{Grebenkov2007}, diffusion on graph \cite{Comtet2002} and diffusion under resetting \cite{PalA2019}.

While the statistical properties of the local time are quite extensively studied for Brownian motion (BM) and its generalisations, very less amount of study has been performed for active processes. We study the statistical properties of the local time for run and tumble particle (RTP) which is an emblematic model for active systems. Such study is particularly relevant in the present day research where the self propelled motion of these particles are harnessed to produce the useful work \cite{Santiago2018} which has the possibility of potential therapeutic application of the active particles as drug delivery machines for diseases like cancer and heart disease \cite{Ghosh2020}. Under this circumstance, one important quantity to keep track of would be the time that the particle spends in the vicinity of the desired location as this time can quantify the efficiency of these machines. Motivated from these applications, we study the local time statistics for the run and tumble particle.

In the random walk literature, run and tumble motion  is known as the persistent Brownian motion and has been substantially studied in the past \cite{Weiss2002, Masoliver2017}. Recently the model has seen surge in interest due to its biological application in modelling the motion of bacteria like E- Coli \cite{Berg2003}. The collective dynamics of interacting RTPs along with their self propelling nature give rise to various novel phenomena like motility induced phase separation \cite{Tailleur2008, Solon2015}, clustering \cite{Sepulveda2016}, non-existence of equation of state  \cite{SolonA2015} and so on. Furthermore, even at the individual level a single RTP exhibits interesting features and  a myriad of its properties are known. Examples include - probability distribution in free space as well as in confining potential \cite{Malakar2018, DemaerelA2018,DharA2019, Basu2020, Majumdarresetting2018}, first passage time properties \cite{Angelani2014, DoussalMaj2019,Mori2019}, arcsine laws \cite{Singh2019}, escape problems \cite{Woillez2019}, convex hull \cite{Hartmann2020}, large deviation forms \cite{Gradenigo2019, Banerjee2019, Santra2020}, distribution of maximum \cite{Masoliver1993,Bertrand2020}, behaviour in inhomogeneous medium \cite{Doussal2020, Singh2020} and so on. Other models like active Brownian particle have also been considerably studied in the recent years \cite{BasuU2018, BasuU2019}. 

In this paper, we investigate the statistical properties of the local time for a RTP in one dimension. Generalising the formalism of Feynman and Kac \cite{Kac1949, Kac1951,Majumdar2005} to the case of RTP, we deploy this method to obtain the distribution and moments of the local time. Furthermore, experimentally it is seen that the motility parameters of E- Coli strongly depend on the nutrient concentration and nutrient gradient. This provides us the impetus to generalise the RTP model in an inhomogeneous environment. Recently in \cite{Singh2020}, such generalisation of RTP model was considered on infinite as well as on semi-infinite line. While the RTP model in homogeneous environment has been substantially studied and a huge amount of results are known, very few results exist for the inhomogeneous RTP model. In an attempt in this direction, we study the statistics of the local time for inhomogeneous RTP model in this paper. As illustrated later, such extensions give rise to non-trivial scaling forms. Moreover in many practical applications, the particle moves in a bounded domain and one is interested in the knowledge of time that the particle spends in the vicinity of some point before getting absorbed by the boundaries \cite{Pal2019}. For example, the enzymatic action of a diffusing protein inside a cell is related to the time that the protein spends inside cell before it finds the correct binding site \cite{Redner2001}. Guided by these applications, we have considered the statistics of local time in semi-infinite line with the absorbing boundary condition in the second part of the paper.

The paper is organised as follows. In sec. \ref{model-int}, we introduce the model and summarise the main results of our paper. Sec. \ref{loc-infinite-line} deals with the local time statistics in an infinite line with sec. \ref{infi-alp-0p0} containing discussions for $\alpha = 0 $ and sec. \ref{gen-alpha} for general $\alpha $. We devote sec. \ref{loc-abs-wall} for the local time in presence of an absorbing wall with sec. \ref{abs-alp} for $\alpha= 0 $ and sec. \ref{abs-alp-gen} for general $\alpha$ followed by the conclusion in sec. \ref{conclusion}.

\section{MODEL AND SUMMARY OF RESULTS}
\label{model-int}
We study the motion of a run and tumble particle (RTP) in one dimension in an inhomogeneous medium. The time evolution equation for the position of the particle is given by
\begin{align}
\frac{dx}{dt} = v \sigma (t),
\label{model-eq-1}
\end{align}
where $v~(>0)$ is the speed of the particle and $\sigma(t)$ is the its instantaneous direction which is governed by the telegraphic or dichotomous noise that alternates between $\pm1$ with some rate $R$. For constant $R$, the noise, at different times, are exponentially correlated as $v^2\langle \sigma (t) \sigma (t') \rangle = v^2e^{-2 R |t-t'|}$ which makes it a non-Markovian process. However in the limit $R \to \infty$, $v \to \infty$ keeping $\frac{v^2}{R}$ fixed, the particle behaves like a Brownian particle with diffusion constant $\frac{v^2}{2 R}$. On the other hand, when $R \to 0$, the particle performs the ballistic motion.

Here we study the generalised version of this model in which the run and tumble particle moves in an inhomogeneous medium. The inhomogeneity is introduced by considering the position dependent rate of flipping $R(x)$ \cite{Singh2020}. In this paper, we focus on the following form of the rate:
\begin{align}
R(x) = \gamma \frac{|x|^{\alpha}}{l^{\alpha}},~~~~~\text{with }\alpha \geq 0,
\label{rate-eq}
\end{align}
where $\gamma$ is a positive constant that sets the timescale for the activity and $l$ is the length over which the rate varies. For $\alpha =0$, we recover the usual RTP model with constant rate $\gamma$.

\begin{figure}[h]
\includegraphics[scale=0.35]{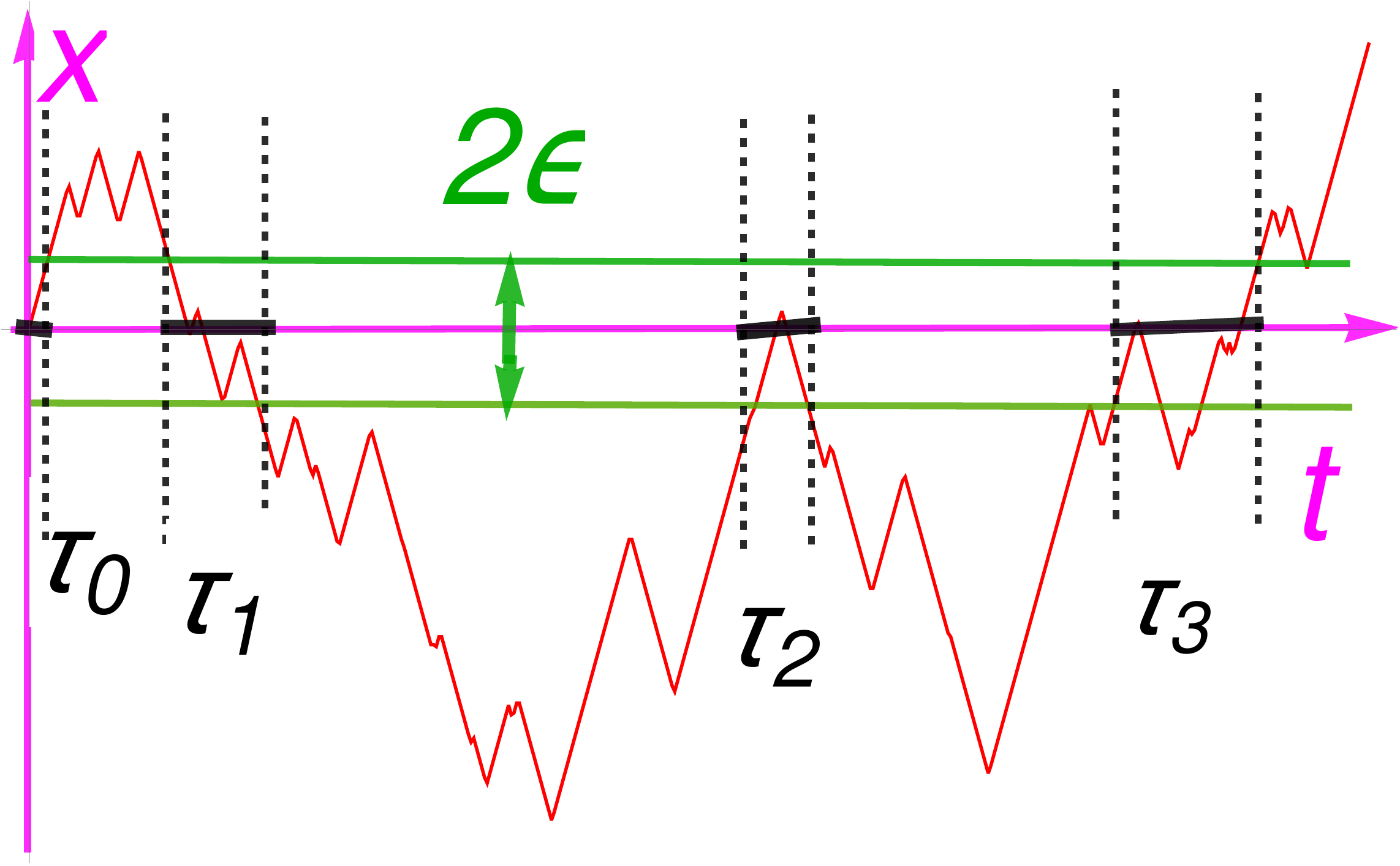}
\centering
\caption{Schematic of the local time for a typical trajectory of the RTP till time $t$ (shown by red line). We consider a small interval of thickness $2\epsilon$ about the origin (shown by green lines). Out of the total time $t$, the particle spends $T_{2 \epsilon}=\tau _0+\tau_1+\tau_2+\tau _3$ amount of time in this interval. For this case, the local time (density) is defined as $T_{loc} = \lim _{\epsilon \to 0}\left[\frac{\tau _0+\tau_1+\tau_2+\tau _3}{2 \epsilon}\right]$.}
\label{trajectory-pic}
\end{figure}

The inhomogeneous model is recently studied in \cite{Singh2020}, where the authors explicitly computed the probability distribution and persistent properties for all values of $\alpha$. Here we focus on the statistical properties of the local time (denisty) which is  defined as 
\begin{align}
T_{loc}(b) = \int _{0}^{t} \delta (x(\tau)-b) d\tau.
\label{model-eq-2}
\end{align} 
Physically $T_{loc}(b) db$ is the time that the particle spends in the region $b$ to $b+db$ out of total time $t$. For simplicity, we take $b=0$ (unless specified) and denote $T_{loc}(0)$ by simply $T_{loc}$. For a typical trajectory of the particle, we have illustrated $T_{loc}$ schematically in Fig. \ref{trajectory-pic}. Using the generalisation of the Feynman-Kac formalism \cite{Kac1949,Kac1951,Majumdar2005}, we compute the probability distribution and moments of $T_{loc}$ for all values of $\alpha$.

In the second part of the paper, we consider the motion of a RTP starting from $x_0=0$ in presence of an absorbing wall at $x=M$ and study the properties of the local time at the origin till it gets absorbed by the wall. We generalise the formalism due to Feynman and Kac and compute the distribution of $T_{loc}$ for all values of $\alpha$. We summarise our results below:

\subsection*{Infinite line}
\begin{itemize}
\item[•]For $\alpha =0$, we derive exactly the moments generating functions from which we provide the expression of first three moments of $T_{loc}$. Using these expressions, we find that the local time scales typically as $T_{loc} \sim \sqrt{t}$ at large $t$ which corresponds to the Brownian limit of the RTP model. On the other hand, as $t \to 0^+$, all moments have non-zero value which is in contrast to the Brownian motion where moments vanish as $t \to 0^+$. 

\item[•]We also compute the exact probability distribution of $T_{loc}$  for $\alpha =0$ in Eq. \eqref{alph-0-dist-eq-3} which consists of a series of appropriately weighted $\delta$-functions as
\begin{align}
&P\left(T_{loc},t\right) = \sum _{m=0}^{\infty} \mathcal{S}_{m}(t)~ \delta \left(T_{loc}-\frac{2m+1}{2v} \right), \label{alph-0-dist-eq-3}\\
&\mathcal{S}_{m}(t) =e^{-\gamma t}\left[I_m(\gamma t) +I_{m+1}(\gamma t) \right].
\label{alph-0-dist-eq-3333}
\end{align}
While at large $t$ and large $T_{loc}$, the distribution $P(T_{loc},t)$ correctly reduces to that of the Brownian motion with diffusion constant $\mathfrak{D}_0 = \frac{v^2}{2 \gamma}$, the short time behaviour is quite  different from that of a Brownian motion. 
For $t<<\frac{1}{\gamma}$, we find that $P(T_{loc},t) \simeq \delta \left(T_{loc}-\frac{1}{2 v} \right)$. 
The coefficient $\mathcal{S}_m(t)$ of the delta function $\delta \left(T_{loc} - \frac{2m+1}{2v}\right)$ can be associated to the probability of $m$-th visit by the RTP to the origin starting from the origin. Thus, as a by-product we also obtain the exact expression for the probability $\mathcal{S}_m(t)$ that the RTP visits the origin $m$-times starting from the origin till time $t$ in Eq. \eqref{alph-0-dist-eq-3333}.

\item[•]For general $\alpha$, we compute the expressions of all moments for large $t$ and they are given by 
\begin{align}
\langle T_{loc}^{n}(t) \rangle  \simeq  \frac{n!}{\mathcal{C}_{\alpha}^n \Gamma \left( \frac{1+(n+1)(1+\alpha)}{2+\alpha}\right)} t^{\frac{n(1+\alpha)}{2+\alpha}},
\label{alph-neq-eq-10}
\end{align}
for $n=1,2,3...$
These late time growths of the moments suggest that the fluctuations of $T_{loc}$ scales typically  as $T_{loc}\sim t^{\frac{1+\alpha}{2+\alpha}}$ with time which is the extension of $\sqrt{t}$ scaling for $\alpha =0$ case.

\item[•]Furthermore, for general $\alpha~(>0)$, we show that at large $t$ the probability distribution for the typical fluctuations of $T_{loc}$ possesses a scaling behaviour of the form 
\begin{align}
P(T_{loc},t) \simeq \frac{\mathcal{C}_{\alpha}}{t^{\frac{1+\alpha}{2+\alpha}}} ~f_{\alpha} \left( \frac{\mathcal{C}_{\alpha} T_{loc}}{t^{\frac{1+\alpha}{2+\alpha}}}\right),
\label{model-eq-3}
\end{align}
where $\mathcal{C}_{\alpha}$ is a constant given in Eq. \eqref{alph-neq-eq-911} and the scaling function $f_{\alpha}(z)$ is given explicitly by
\begin{align}
f_{\alpha}(z) = \frac{1}{\pi}\sum _{n=1}^{\infty} &\frac{(-z)^{n-1}}{n! } \frac{\Gamma\left( 1+n \frac{1+\alpha}{2+\alpha}\right)}{\left(  \frac{1+\alpha}{2+\alpha}\right)}  \nonumber \\  
&\times~
\sin \left(\pi n  \frac{1+\alpha}{2+\alpha} \right).
\label{alph-neq-eq-13}
\end{align}
Using saddle point approximation, we show that $f_{\alpha}(z)$ for large $z$ decays as \\ $~~~~~~~~~f_{\alpha}(z) \sim  \text{exp}\left( -\frac{(1+\alpha)^{1+\alpha}}{(2+\alpha)^{2+\alpha}}~ z^{2+\alpha}\right).$ 

\end{itemize}  
\subsection*{In presence of an absorbing wall}
\begin{itemize}
\item[•]In the second part of the paper, we look at the distribution of $T_{loc}$ in presence of an absorbing wall at $x=M~(>0)$. Here $T_{loc}$ is defined as the local time (density) about the origin of the RTP starting from $x_0 = \epsilon ~(\to 0^+)$ before getting absorbed by the wall. 
For $\alpha = 0$, by solving the backward equations exactly, we obtain the distribution  $\mathcal{P}_{\pm}(T_{loc})$ of $T_{loc}$ and find that the distribution in this case also consists of a series of $\delta$-functions as 
\begin{align}
&~~~~~~\mathcal{P}_+(T_{loc})= \sum _{m=0}^{\infty} \mathfrak{p}_{0} \left( 1-\mathfrak{p}_{0} \right)^{m} \delta \left( T_{loc}-\frac{2 m}{v}\right), \label{abs-0-eq-10}\\
&~~~~~~\mathcal{P}_-(T_{loc})= \sum _{m=1}^{\infty} \frac{\mathfrak{p}_{0}\left( 1-\mathfrak{p}_{0} \right)^{m}}{( 1-\mathfrak{p}_{0} )}  \delta \left( T_{loc}-\frac{2 m}{v}\right),
\label{abs-0-eq-11}
\end{align} 
where $\mathfrak{p}_0 = \frac{v}{v+\gamma M}$ and subscript ``$ \pm "$ in $\mathcal{P}_{\pm}(T_{loc})$ denotes the initial velocity direction. Unlike in the infinite line case, we find that the distribution can be sensitive to the initial velocity depending on the initial position.
Once again, based on path-counting analysis, we argue that the $\delta \left(T_{loc}-\frac{2m}{v}\right)$ arises due to the fact that the RTP crosses the origin $m$-times before getting absorbed by the wall. The coefficient of $\delta \left(T_{loc}-\frac{2m}{v}\right)$ in Eqs. \eqref{abs-0-eq-10} and \eqref{abs-0-eq-11} is just the probability that the RTP visits the origin $m$-times starting from the origin before getting absorbed at the wall.

\item[•]It turns out that one can extend the the path-counting analysis (done for $\alpha =0$)  to general $\alpha>0$ which guides us to generalise $\mathcal{P}_{\pm}(T_{loc})$ in Eqs. \eqref{abs-0-eq-10} and \eqref{abs-0-eq-11} for general $\alpha >0$ as
\begin{align}
&~~~~\mathcal{P}_+(T_{loc})= \sum _{m=0}^{\infty} \mathfrak{p}_{\alpha} \left( 1-\mathfrak{p}_{\alpha} \right)^{m} \delta \left( T_{loc}-\frac{2 m}{v}\right), \label{abs-n0-eq-1}\\
&~~~~\mathcal{P}_-(T_{loc})= \sum _{m=1}^{\infty}  \frac{\mathfrak{p}_{\alpha}\left( 1-\mathfrak{p}_{\alpha} \right)^{m}}{( 1-\mathfrak{p}_{\alpha} )} \delta \left( T_{loc}-\frac{2 m}{v}\right),
\label{abs-n0-eq-2}
\end{align} 
where $\mathfrak{p}_{\alpha} = \frac{v (1+\alpha)l^{\alpha}}{\gamma M^{1+\alpha}+v (1+\alpha)l^{\alpha}}$ represents the probability that the particle starting from $0^+$ gets absorbed at the absorbing boundary without ever revisiting the origin. 
\end{itemize}
In what follows, we provide a detailed derivation of these results. For the clarity of presentation, we will relegate some calculations to the appendices.

\begin{figure*}[t]
  \centering
  \subfigure{\includegraphics[scale=0.22]{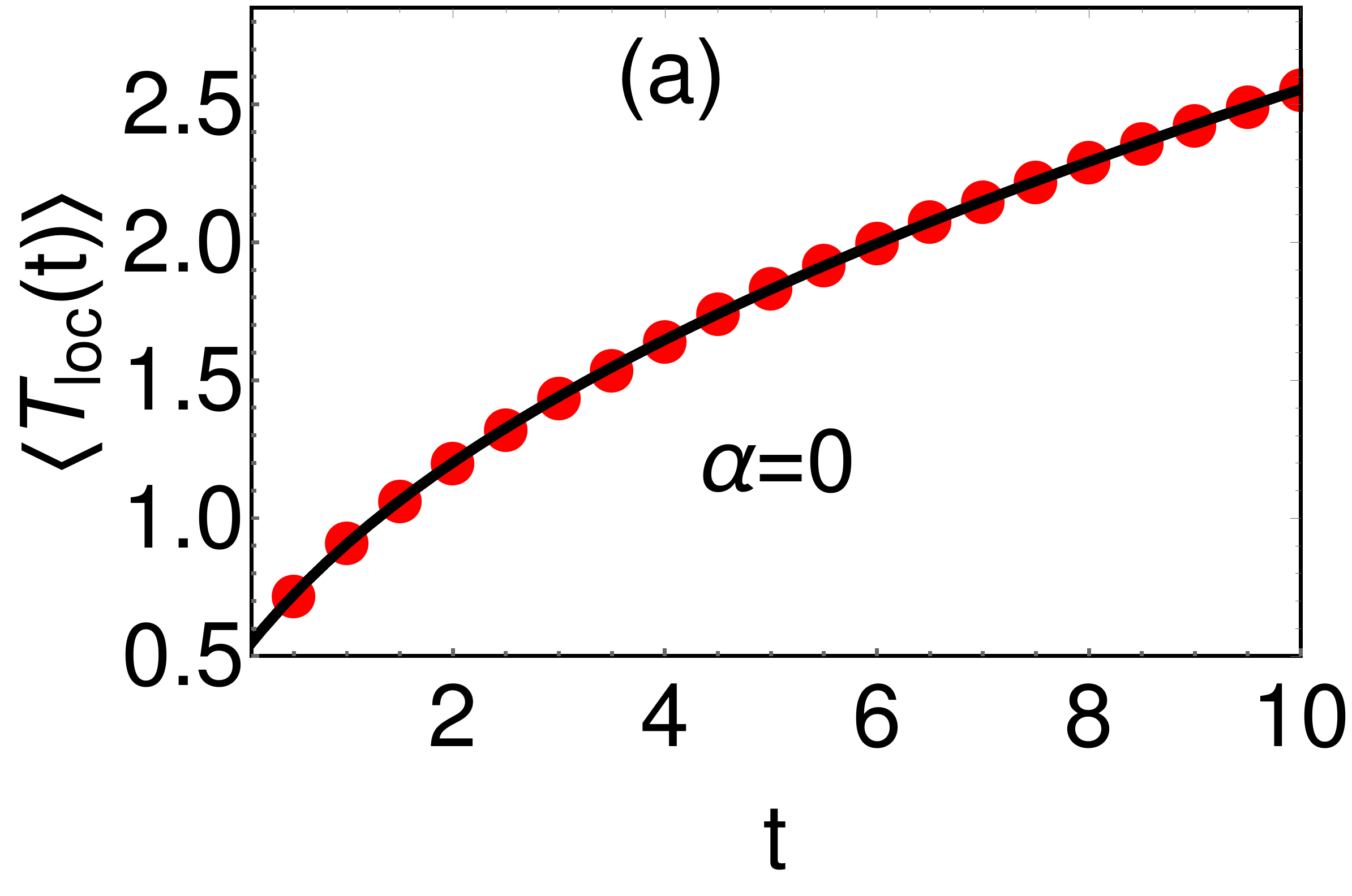}}
  \subfigure{\includegraphics[scale=0.22]{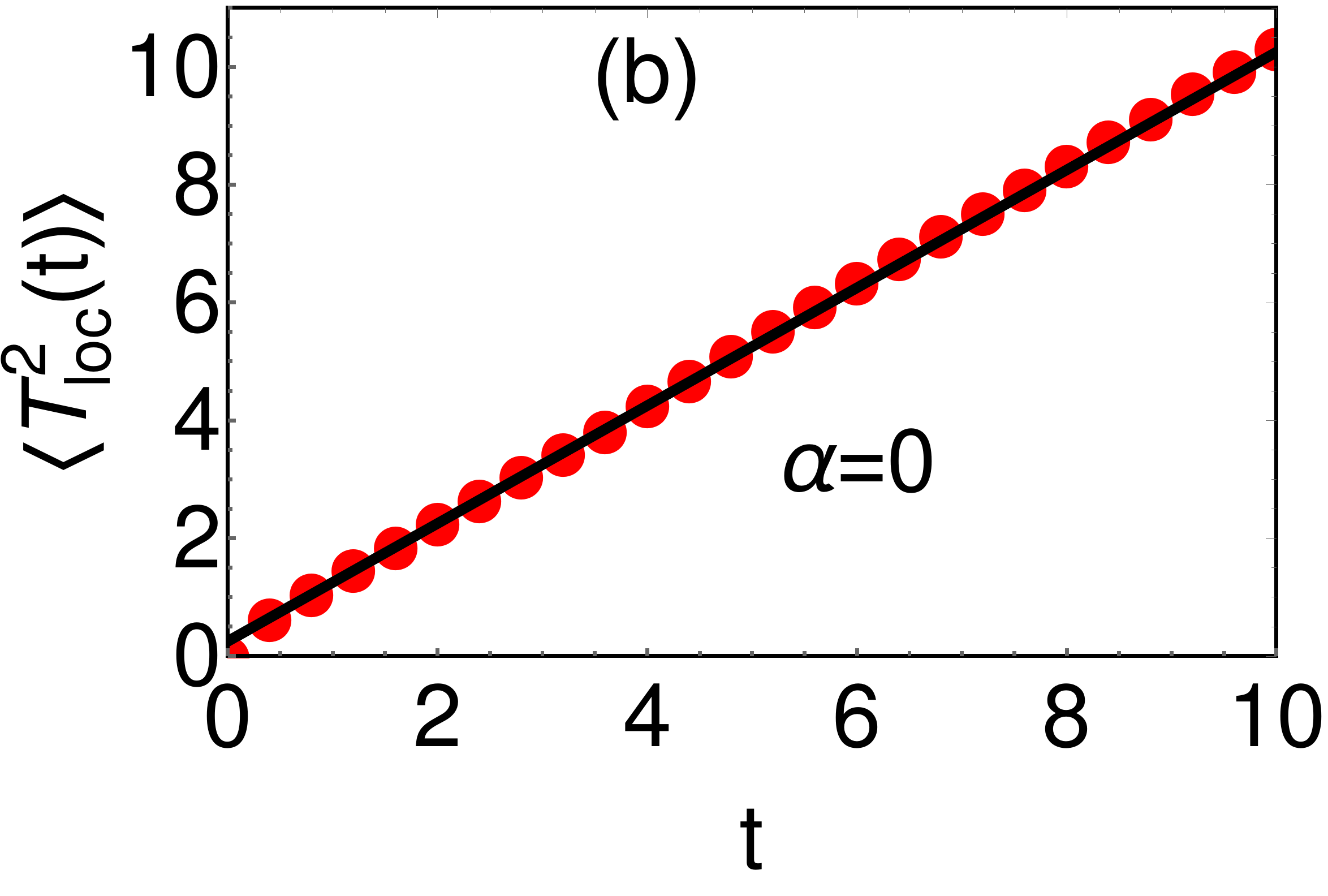}}
  \subfigure{\includegraphics[scale=0.22]{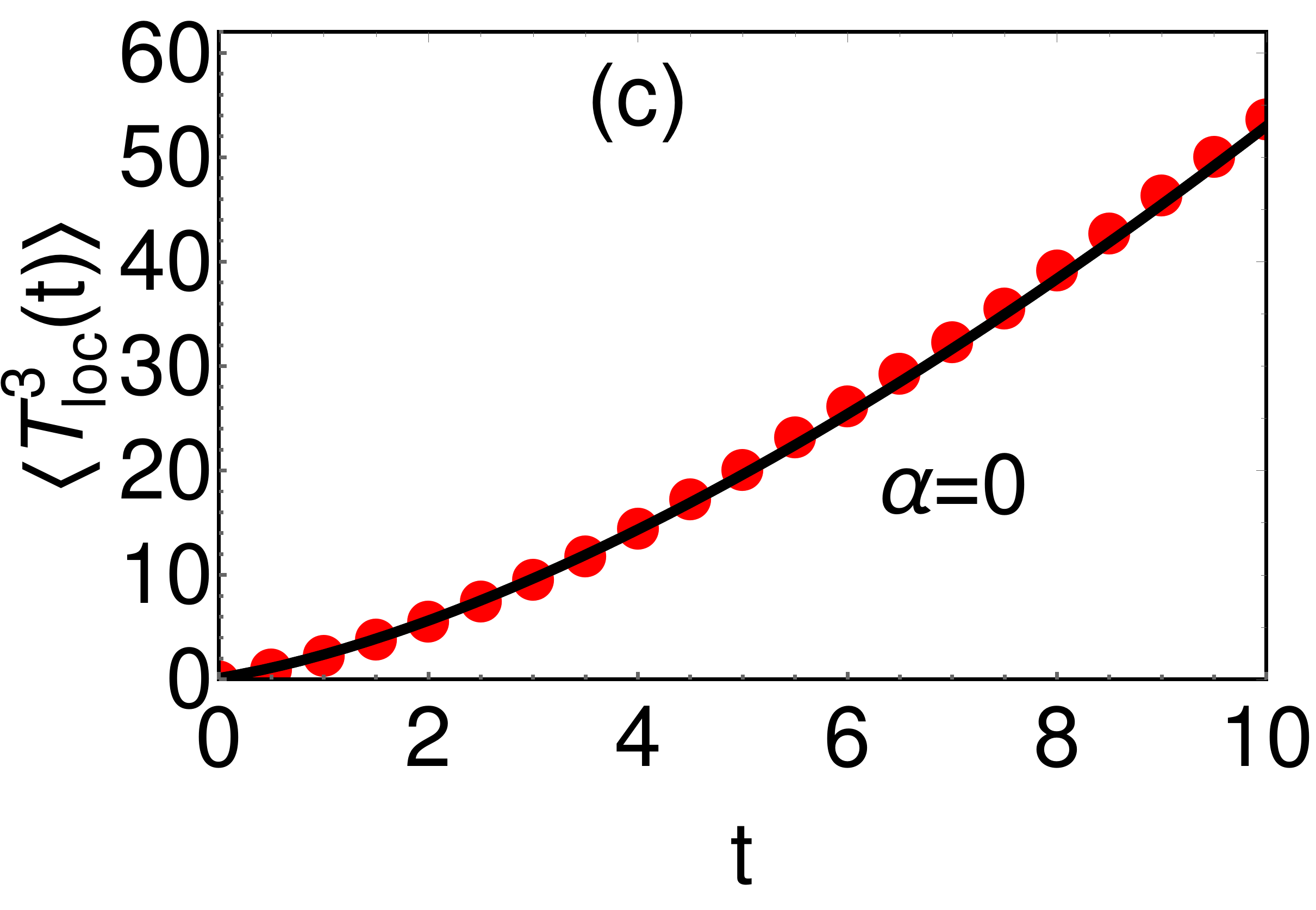}}
  \caption{Plot of the first three moments of $T_{loc}$ (shown by solid line) as a function of $t$ for $\alpha =0$ as obtained in Eq. \eqref{new-mom-alph-0p0} and comparision with the numerical simulations (shown by red filled circles). We have taken $v=1,~\gamma=1$. The simulation is conducted with $\epsilon =0.001$. for all plots. }
\label{alp-0p0-mom}  
\end{figure*}

\section{LOCAL TIME STATISTICS IN AN INFINITE LINE}
\label{loc-infinite-line}
We begin with the calculations of the moments and distribution of $T_{loc}$ for inhomogeneous RTP model with rate $R(x)$ defined in Eq. \eqref{rate-eq}. We denote the probability distribution of $T_{loc}$ as $P_{\pm}\left(T_{loc},x_0,t \right)$ given that the particle initially starts from $x_0$ with velocity $\pm v$ and is allowed to evolve till time $t$. We define the Laplace transformation of $P_{\pm}\left(T_{loc},x_0,t \right)$ with respect to $T_{loc}$ as
\begin{align}
Q_{\pm} (p,x_0,t) = \int _{0}^{\infty} ~dT_{loc} ~e^{-p T_{loc}}~P_{\pm}\left(T_{loc},x_0,t \right),
\label{def-qpm}
\end{align}
where $p$ is the conjugate variable of $T_{loc}$.\\
To derive the statistics of $T_{loc}$, we proceed as follows: Intitively, the local time (density) can be understood, following P. L\'evy \cite{Balkema1990}, as 
\begin{align}
&T_{loc} = \lim _{\epsilon \to 0^+} \frac{T_{2 \epsilon}}{2 \epsilon},~\text{with }\label{new-infi-eq-1}\\
& T_{2 \epsilon} = \int _{0}^{t} \mathbb{I} _{\epsilon} \left( x(\tau)\right)  d\tau,
\label{new-infi-eq-2}
\end{align}
where $\mathbb{I} _{\epsilon} \left( x(\tau)\right)=\Theta \left( x(\tau)+\epsilon \right) \Theta \left( \epsilon-x(\tau) \right)$ with $\Theta (x)$ being the Heaviside Theta function and $\epsilon >0$. The choice of the Heaviside function ensures that $\mathbb{I} _{\epsilon} \left( x\right)$ is $1$ for $-\epsilon < x <\epsilon$ and $0$ otherwise. Here $T_{2 \epsilon}$ is the time spent by the RTP inside a small but non-zero interval $x \in [-\epsilon, \epsilon]$ till time $t$. Denoting the distribution of $T_{2 \epsilon}$ by $G_{\pm}(T_{2 \epsilon}, x_0, t)$ with initial velocity $\pm v$, we consider the Laplace transform with respect to $T_{2 \epsilon}$ as 
\begin{align}
H_{\pm} (q,x_0,t) = \int _{0}^{\infty} ~dT_{2 \epsilon} ~e^{-q T_{2 \epsilon}}~G_{\pm}\left(T_{2 \epsilon},x_0,t \right).
\label{new-infi-eq-4}
\end{align}
Following the formalism due to Feynman and Kac \cite{Kac1949,Kac1951,Majumdar2005}, we write the backward master equations for $H_{\pm} (p,x_0,t)$:
\begin{align}
\partial _t H_+ &=~~ v \partial _{x_0} H_+ -R(x_0)H_+ +R(x_0) H_- -q \mathbb{I} _{\epsilon}\left( x_0\right) H_+, \nonumber \\
\partial _t H_- &= -v \partial _{x_0} H_- +R(x_0)H_+ -R(x_0) H_- -q \mathbb{I} _{\epsilon}\left( x_0\right)  H_-,
\label{new-infi-eq-5}
\end{align}
with $R(x)$ defined in Eq. \eqref{rate-eq}. These equations are derived explicitly in Appendix \ref{BME-Q}. Our aim now is to solve these coupled equations for $\alpha \geq 0$. To solve them, we have to specify the initial condition and the boundary conditions which are given by   
\begin{align}
& H_{\pm}(q, x_0, t=0)=1, \label{ini-eq-1} \\
& H_{\pm}(q, x_0 \to \pm \infty, t)=1.
\label{bcs-eq-1}
\end{align}
 To understand Eq. \eqref{ini-eq-1}, we note from the definition of $T_{ 2 \epsilon}$ in Eq. \eqref{new-infi-eq-2} that, $T_{ 2 \epsilon} \to 0$ in the limit $t \to 0$, which in turn provides $H_{\pm} (q,x_0,t \to 0) = 1$.
The boundary conditions in Eq. \eqref{bcs-eq-1} can be understood in the following way. If $x_0 \to \pm \infty$, then the particle initially is very far away from the origin and it will not reach the interval $x \in [-\epsilon, \epsilon]$ in any finite time $t$. This implies that the particle does not spend any time in the interval which results in $T_{ 2 \epsilon} $ being equal to zero. In other words, $G_{\pm}(T_{ 2 \epsilon},x_0 \to \pm \infty,t) = \delta (T_{ 2 \epsilon})$. Inserting this form of $G_{\pm}(T_{ 2 \epsilon},x_0 \to \pm \infty,t)$ in Eq. \eqref{new-infi-eq-4}, we see that $H_{\pm} (q,x_0 \to \pm \infty,t) = 1$. \\
Recall that ultimately, the aim is to compute $P_{\pm}\left(T_{loc},x_0,t \right)$. To obtain this distribution, we substitute the definition of $T_{loc}$ from Eq. \eqref{new-infi-eq-1} in $Q_{\pm}\left( p,x_0,t\right)$ in Eq. \eqref{def-qpm} and use $P_{\pm}\left(T_{loc},x_0,t \right) dT_{loc} = G_{\pm}(T_{2 \epsilon}, x_0, t)dT_{2 \epsilon}$ which yields
\begin{align}
Q_{\pm}\left( p,x_0,t\right) = \lim _{\epsilon \to 0}H_{\pm}\left( \frac{p}{2 \epsilon},x_0,t\right).
\label{new-infi-eq-6}
\end{align}
In what follows, we solve the backward equations \eqref{new-infi-eq-5} with the conditions in Eqs. \eqref{ini-eq-1} and \eqref{bcs-eq-1} and then use Eq. \eqref{new-infi-eq-6} to compute the distribution of $T_{loc}$. To solve Eqs. \eqref{new-infi-eq-5}, we take the Laplace transformation of $H_{\pm} (p,x_0,t)$ with respect to $t$ as
\begin{align}
\bar{H}_{\pm} (q,x_0,s) = \int _{0}^{\infty}dt~e^{-s t}~H_{\pm} (q,x_0,t).
\label{LT-eq-1}
\end{align}
It is straightforward to translate the boundary conditions in Eq. \eqref{bcs-eq-1} in terms of $\bar{H}_\pm(q,x_0,s)$ as 
\begin{align}
\bar{H}_\pm(q,x_0 \to \pm \infty,s) = \frac{1}{s}. 
\label{barQ-s-bc}
\end{align} 
Introducing the following two new functions
\begin{align}
&2 \bar{H}(q,x_0,s) =\bar{H}_+(q,x_0,s) +\bar{H}_-(q,x_0,s), \label{Q-Z-eq-1}\\ 
&2 \bar{Z}(q,x_0,s) =\bar{H}_+(q,x_0,s) -\bar{H}_-(q,x_0,s),
\label{Q-Z-eq-2}
\end{align}
we rewrite the master equations \eqref{new-infi-eq-5} in Laplace space as
\begin{align}
&s \bar{H}-1 = v \partial _{x_0} \bar{Z} - q \mathbb{I}_{\epsilon} \left(  x_0 \right) \bar{H}, \label{bFP-eq-Lap-2}\\
&s \bar{Z} = v \partial _{x_0} \bar{H} -2 R(x_0)\bar{Z}- q \mathbb{I}_{\epsilon} \left(  x_0 \right) \bar{Z}. 
\label{bFP-eq-Lap-3}
\end{align}
The second equation can be used to write $\bar{Z}(q,x_0,s)$ in terms of $\bar{H}(q,x_0,s)$ which then can be substituted in Eq. \eqref{bFP-eq-Lap-2} to get a differential equation only for $\bar{H}(q,x_0,s)$ which reads
\begin{align}
s \bar{H}-1 =  \frac{\partial}{\partial x_0} \left(\frac{v^2}{s+2 R(x_0)+q \mathbb{I}_{\epsilon}} \frac{\partial \bar{H}}{\partial x_0}\right) - q \mathbb{I}_{\epsilon} \left(  x_0 \right) \bar{H}.
\label{Q-eq-1}
\end{align}
In what follows, we will first consider the case $\alpha =0$ for which we are able to solve it exactly. From this solution, we obtain the moments and distribution of $T_{loc}$. In the subsequent section \ref{gen-alpha} we consider the general $\alpha$ case for which we provide approximate solution that is valid for small $s$ (equivalently large $t$).
\begin{figure*}[t]
  \centering
  \subfigure{\includegraphics[scale=0.3]{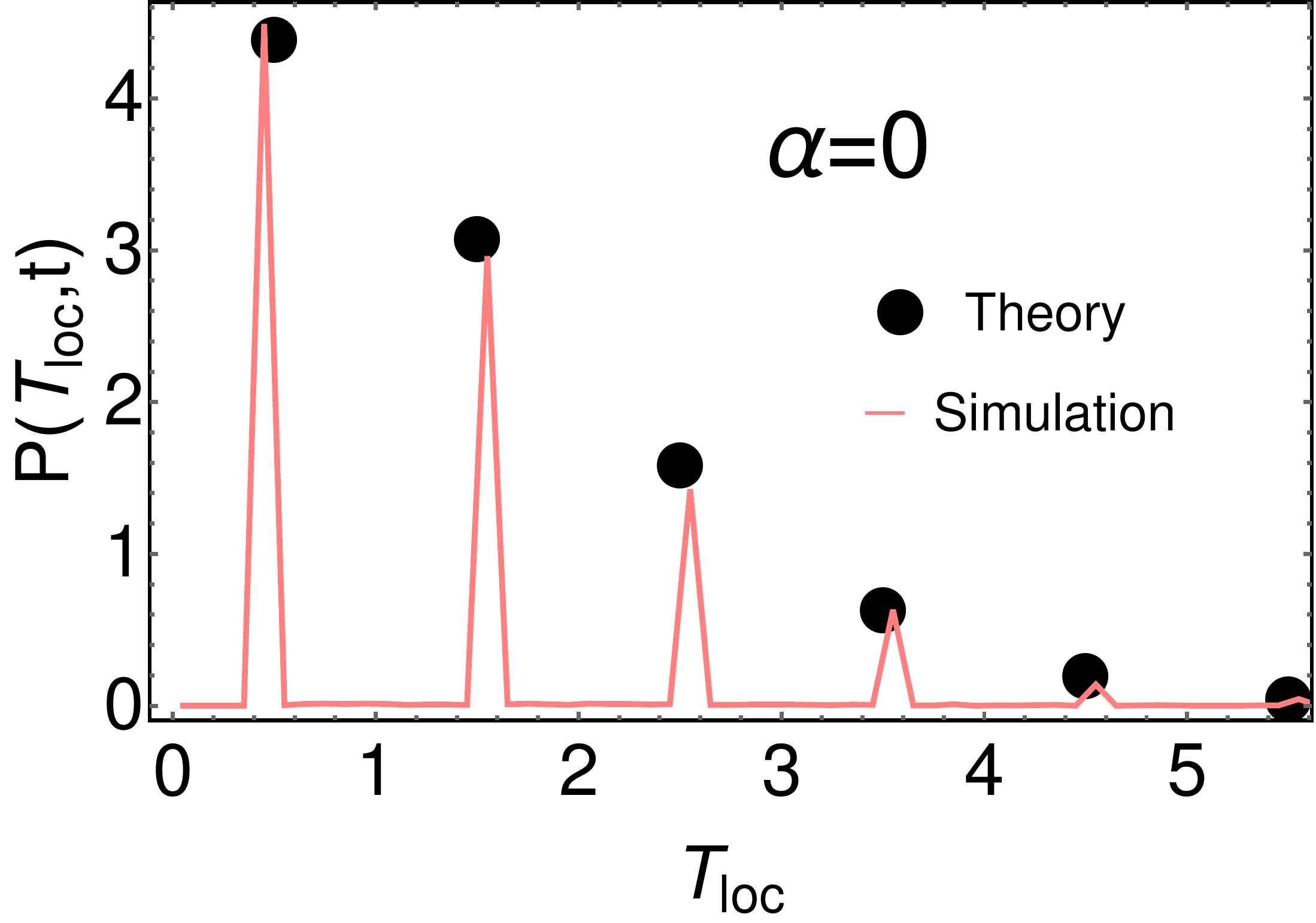}}
  \subfigure{\includegraphics[scale=0.3]{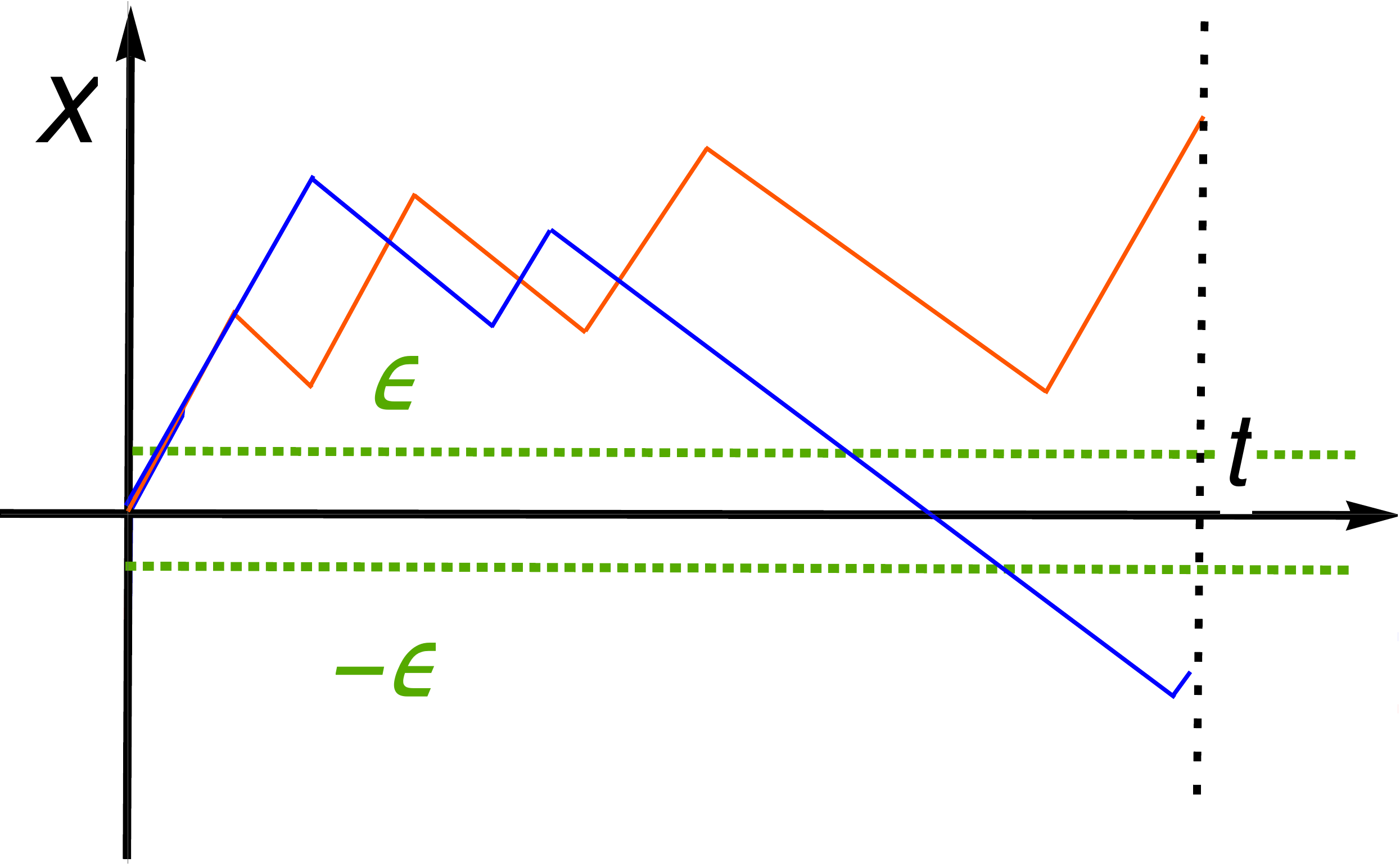}}
  \caption{\textit{Left panel:} Numerical verification of the local time distribution $P(T_{loc}, t)$  in Eq. \eqref{alph-0-dist-eq-3} for $\alpha =0$ for $v=1,~\gamma=1$ and $t=3$. For simulation, we have chosen $\epsilon =0.01$. \textit{Right panel:} Schematic of the realisations that give rise to various terms in $P(T_{loc},t)$ in Eq. \eqref{alph-0-dist-eq-3}. For the red trajectory, starting from the origin, the RTP does not cross the origin till time $t$ and contributes $T_{loc} = \frac{\epsilon}{2 \epsilon v}=\frac{1}{2 v}$ to $P(T_{loc}, t)$. Similarly, for blue trajectory, the particle crosses the origin once till time $t$ and contributes $T_{loc} = \frac{3\epsilon}{2 \epsilon v}=\frac{3}{2 v}$ to $P(T_{loc}, t)$.}
\label{prob-alph-0p0-pic-1}  
\end{figure*}
\subsection{Case I: $\alpha =0$}
\label{infi-alp-0p0}
For $\alpha =0$, the rate function $R(x_0)$ becomes $x_0$ independent and is given by a constant value $R(x_0) = \gamma$. This corresponds to the RTP model in a homogeneous medium \cite{Malakar2018}. For this case, Eq. \eqref{Q-eq-1} can be solved exactly for all values of $s$ and $q$. Rewriting this equation, we have
\begin{align}
s \bar{H}-1 =  \frac{\partial}{\partial x_0} \left(\frac{v^2}{s+2 \gamma+q \mathbb{I}_{\epsilon}} \frac{\partial \bar{H}}{\partial x_0}\right) - q \mathbb{I}_{\epsilon} \left(  x_0 \right) \bar{H}.
\label{alph-0-Q-eq-1}
\end{align}
It is easy to solve this equation along with the  boundary conditions in Eq.~\eqref{barQ-s-bc}. For clarity of the presentation, we have consigned the details of the derivation to Appendix \ref{sol-alph-0p0} and present only the end result here. The final expressions for $\bar{H}_{\pm}\left(q,x_0,s \right)$ read
\begin{widetext}
\begin{align}
\bar{H}_+\left(q,x_0,s \right)=
\begin{cases}
\frac{1}{s} + \mathcal{A}_1(q,s) e^{\frac{\lambda_s x_0}{v}}, ~~~~~~~~~~~~~~~~~~~~~~~~~~~~~~~~~~~~~~~~~~~~~~~~~~~\text{if }-\infty <x_0 <-\epsilon,       \\
\frac{1}{s+q} + \mathcal{A}_2(q,s)e^{\frac{\lambda_q x_0}{v}}+\mathcal{A}_2(q,s)\left( \frac{-\lambda _q +\gamma +s+q}{\gamma}\right)e^{\frac{-\lambda_q x_0}{v}},~~~~~~\text{if }-\epsilon <x_0 <\epsilon,  \\
\frac{1}{s} + \mathcal{A}_1(q,s) \left(\frac{-\lambda _s +\gamma+s}{\gamma} \right)e^{\frac{-\lambda_s x_0}{v}},~~~~~~~~~~~~~~~~~~~~~~~~~~~~~~~~~\text{if } \epsilon <x_0 <\infty.
\end{cases}
\label{new-alp-0p0-eq-2}
\end{align}
\begin{align}
\bar{H}_-\left(q,x_0,s \right)=
\begin{cases}
\frac{1}{s} + \mathcal{A}_1(q,s) \left(\frac{-\lambda _s +\gamma+s}{\gamma} \right)e^{\frac{\lambda_s x_0}{v}},~~~~~~~~~~~~~~~~~~~~~~~~~~~~~~~~~\text{if } -\infty <x_0 <-\epsilon,     \\
\frac{1}{s+q} + \mathcal{A}_2(q,s)e^{-\frac{\lambda_q x_0}{v}}+\mathcal{A}_2(q,s)\left( \frac{-\lambda _q +\gamma +s+q}{\gamma}\right)e^{\frac{\lambda_q x_0}{v}},~~~~\text{if }-\epsilon <x_0 <\epsilon,  \\
\frac{1}{s} + \mathcal{A}_1(q,s) e^{-\frac{\lambda_s x_0}{v}}, ~~~~~~~~~~~~~~~~~~~~~~~~~~~~~~~~~~~~~~~~~~~~~~~\text{if }\epsilon <x_0 <\infty,
\end{cases}
\label{new-alp-0p0-eq-3}
\end{align} 
\end{widetext}
where $\lambda _s = \sqrt{s(s+2 \gamma)}$, $\lambda _q = \sqrt{(s+q)(s+q+2 \gamma)}$ and $\mathcal{A}_1(q,s)$ and $\mathcal{A}_2(q,s)$ are $x_0$ - independent functions that need to be computed. Note that the solutions $\bar{H}_{\pm}\left(q,x_0,s \right)$ satisfy the symmetry $\bar{H}_+\left(q,-x_0,s \right)=\bar{H}_-\left(q,x_0,s \right)$. To evaluate the functions $\mathcal{A}_1(q,s)$ and $\mathcal{A}_2(q,s)$, we use the continuity of these solutions at $x_0 = \pm \epsilon$ which gives rise to two linear equations for $\mathcal{A}_1(q,s)$ and $\mathcal{A}_2(q,s)$. Solving these equations provides $\mathcal{A}_1(q,s)$ and $\mathcal{A}_2(q,s)$ which can then be used to compute the distribution of residence time $T_{2 \epsilon}$ in the interval $x \in [-\epsilon, \epsilon]$. However in this paper, since we are interested in computing the distribution of the local time $T_{loc}$ via Eq. \eqref{new-infi-eq-6}, we will evaluate these functions at $q = \frac{p}{2 \epsilon}$ and take $\epsilon \to 0$. Also, for simplicity, we choose $x_0 = 0$ and provide only the expression of $\mathcal{A}_2 \left(\frac{p}{2 \epsilon} ,s\right)$ as
\begin{align}
\lim _{\epsilon \to 0}\mathcal{A}_2 \left(\frac{p}{2 \epsilon} ,s\right) = \frac{\lambda _s +s}{s \left(\lambda _s +s +\gamma - \gamma e^{-\frac{p}{v}} \right)} e^{-\frac{p}{2 v}}.
\label{new-alp-0p0-eq-4}
\end{align}
Inserting this expression in the middle equation of Eqs. \eqref{new-alp-0p0-eq-2}, we get the expression of $\bar{\mathcal{H}}_{+}\left( \frac{p}{2 \epsilon},0,s\right)$ which by symmetry is also equal to $\bar{\mathcal{H}}_{-}\left( \frac{p}{2 \epsilon},0,s\right)$. This can also be verified by inserting $\lim _{\epsilon \to 0}\mathcal{A}_2 \left(\frac{p}{2 \epsilon} ,s\right)$ in $\bar{\mathcal{H}}_{-}\left( \frac{p}{2 \epsilon},0,s\right)$ in the middle equation of Eqs. \eqref{new-alp-0p0-eq-3}. Finally, taking Laplace transformation on both sides of Eq. \eqref{new-infi-eq-6}, we get $\bar{Q}_{\pm }(p,0,s) = \lim_{\epsilon \to 0} \bar{\mathcal{H}}_{\pm}\left( \frac{p}{2 \epsilon},0,s\right)$ using which we get
\begin{align}
\bar{Q}(p,s) =\frac{(\lambda _s +s)e^{-\frac{p}{2 v}}}{s \left(\lambda _s +s +\gamma - \gamma e^{-\frac{p}{v}} \right)}, 
\label{alph-0-eq-9}
\end{align}
where we have used the notation $\bar{Q}(p,s)$ instead of $\bar{Q}_{\pm }(p,0,s)$ since it is symmetric with respect to the initial velocity because $x_0=0$. Hence, we also drop the subscripts $\pm$ in the distribution $P_\pm(T_{loc},t)$ and denote it simply by  $P(T_{loc},t)$. To get the distribution $P(T_{loc},t)$, one has to perform two inverse Laplace transformations - one with respect to $s$ and the other with respect to $p$. We can also compute the moments of $T_{loc}$ by appropriately differentiating $\bar{Q}(p,s)$ with respect to $p$. In the next, we first calculate the moments of $T_{loc}$ followed by the derivation of the distribution $P(T_{loc},t)$.
\begin{figure*}[t]
  \centering
  \subfigure{\includegraphics[scale=0.2]{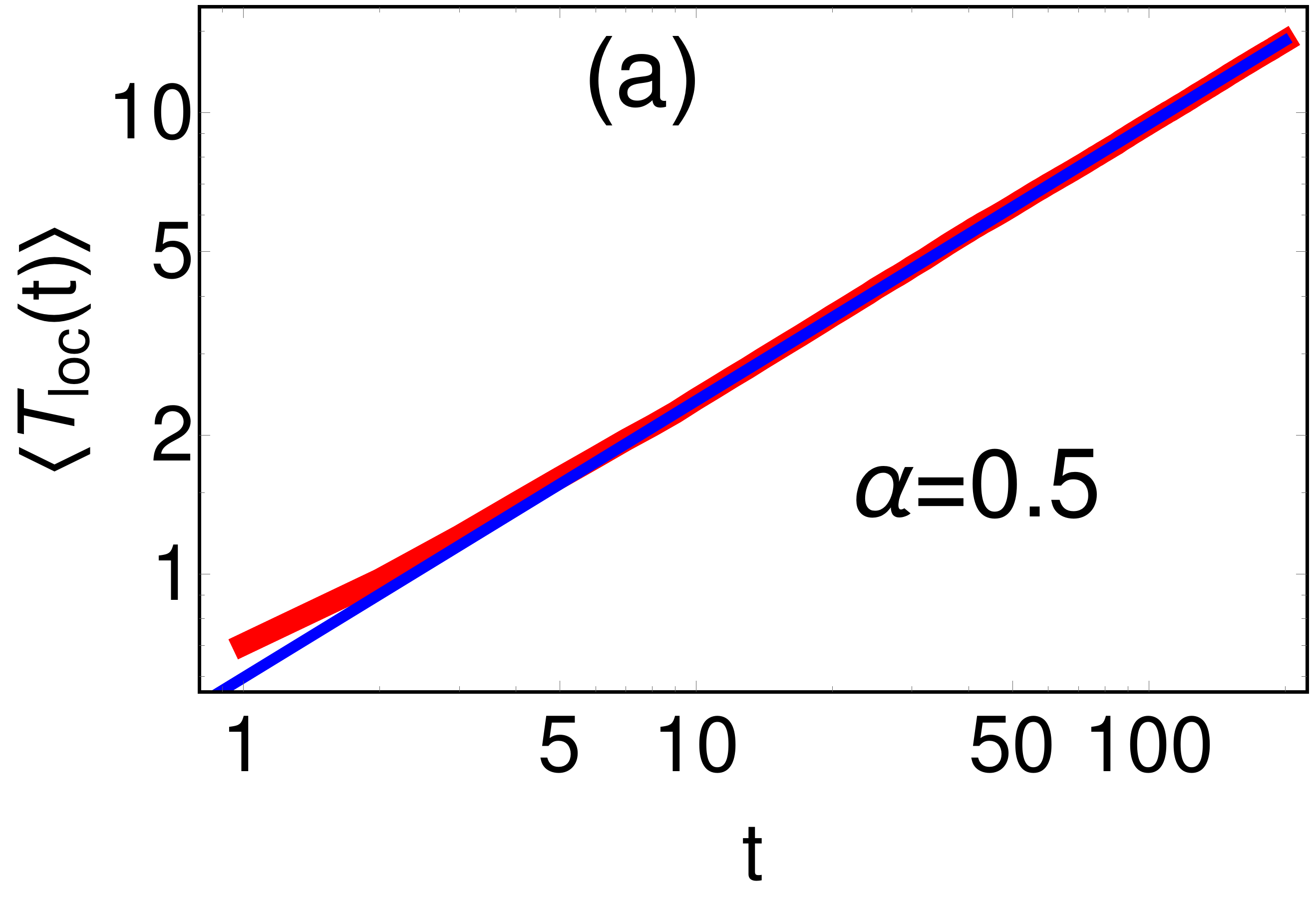}}
  \subfigure{\includegraphics[scale=0.21]{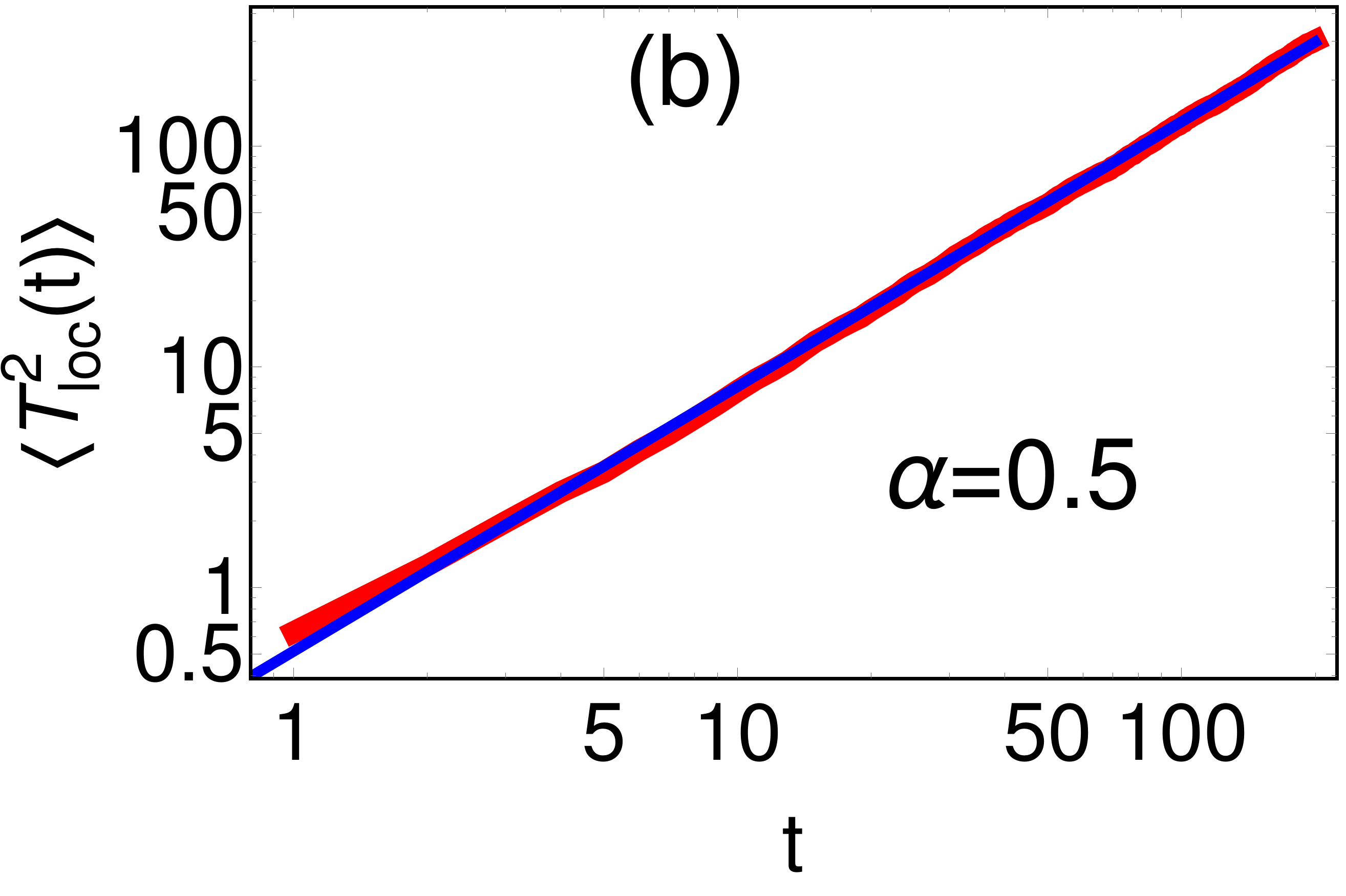}}
  \subfigure{\includegraphics[scale=0.21]{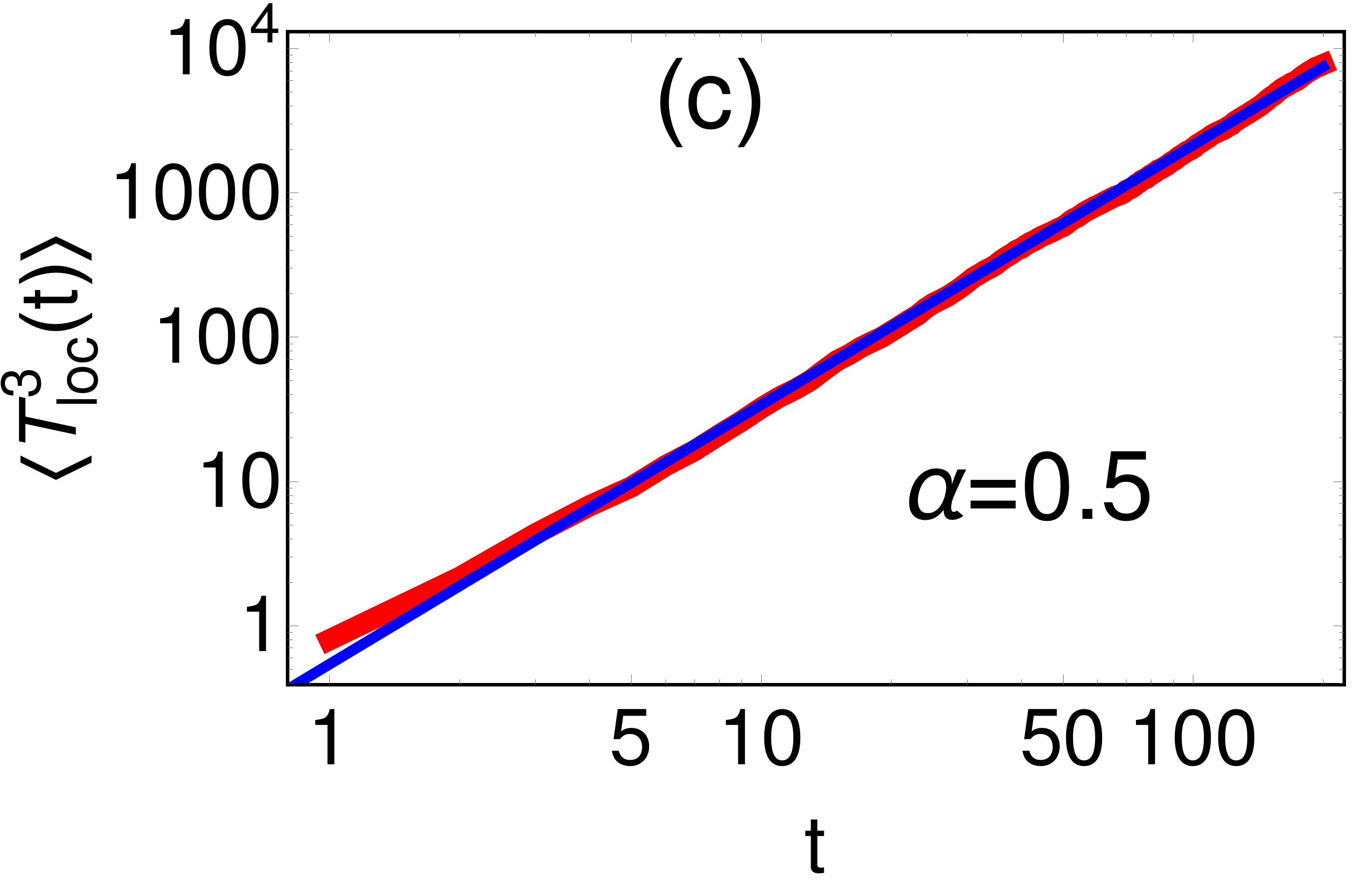}}
  \caption{Plot of different moments of $T_{loc}$ (shown by red line) as a function of $t$ for $\alpha = 0.5$ as obtained in Eq. \eqref{alph-neq-eq-10} and comparision with the numerical simulations (shown by blue line). Values of oher parameters chosen are $v=1,~\gamma=1,~l=1$. The slopes of the plots are (a) $\frac{3}{5}$, (b) $\frac{6}{5}$ and (c) $\frac{9}{5}$.}
\label{alp-neq-mom}  
\end{figure*}
\subsubsection{Moments $\langle T_{loc}^{n}(t) \rangle$ for $\alpha =0$}
Let us first look at the moments of $T_{loc}$ for $\alpha =0$. The moments of $T_{loc}$ are obtained from the derivatives of $\bar{Q}(p,s)$ with respect to $p$ as  
\begin{align}
\langle T_{loc}^{n}(t) \rangle = \mathcal{I}_{s \to t}^{-1} \left[ (-1)^n \left( \frac{\partial ^n \bar{Q}(p,s)}{\partial p^n}\right)_{p=0}\right],
\label{alph-0-mom-eq-1}
\end{align}
where $\mathcal{I}_{s \to t}^{-1}$ indicates the inverse Laplace trasformation from $s \to t$ and is defined for a general function $g(s)$ in terms of the Bromwich integral as
\begin{align}
\mathcal{I}_{s \to t}^{-1} \left[g(s) \right] =  \int_{E -i \infty}^{E+i \infty}\frac{ds}{2 \pi i} ~e^{s t} ~g(s),
\label{inverse-lap}
\end{align}
where $E$ is chosen in such a way that all the singularities are on the left of the Bromwich contour. Inserting $\bar{Q}(p,s)$ from Eq. \eqref{alph-0-eq-9} in Eq. \eqref{alph-0-mom-eq-1}, one can obtain the expressions of the individual moments although getting a closed form expression for the $n$-th order moment turns out to be difficult. Below, we provide the exact expression of the first three moments of $T_{loc}$:
\begin{align}
&\langle T_{loc}(t) \rangle = \frac{e^{-\gamma t}}{2 v} \left[(1+2 \gamma t)I_0(\gamma t) + 2 \gamma t I_1(\gamma t)\right], \nonumber \\
&\langle T_{loc}^2(t) \rangle = \frac{1}{4 v^2} \left( 1+4 \gamma t\right), \nonumber \\
&\langle T_{loc}^3 (t)\rangle = \frac{e^{-\gamma t}}{8v^3} \left[(1+14 \gamma t+16 \gamma ^2 t^2) I_0(\gamma t) \right. \nonumber \\
& \left.~~~~~~~~~~~~~~~~~~~~~~~~+ \left(6 \gamma t + 16 \gamma ^2 t^2 \right) I_1(\gamma t) \right],
\label{new-mom-alph-0p0}
\end{align}
where   $I_{\nu}(\gamma t)$ stands for the modified Bessel function of first kind. In Fig.\ref{alp-0p0-mom}, we have plotted the first three moments of $T_{loc}$ and compared them against the same obtained from numerical simulations. We observe excellent  agreement for all of them.\\
To contrast these expressions with that of the Brownian motion, it is instructive to look at the asymptotic forms of $\langle T_{loc}^{n}(t) \rangle $ for small $t$. For small $t$ (or equivalently large $s$), one gets $\lambda _s = \sqrt{s(2 \gamma +s)} \simeq s$ which when substituted in $\bar{Q}(p,s)$ in Eq. \eqref{alph-0-eq-9} gives $\bar{Q}(p,s) \simeq \frac{e^{-\frac{p}{2 v}}}{s}$. Inserting this expression in Eq. \eqref{alph-0-mom-eq-1} gives the moments for small $t$ as
\begin{align}
\langle T_{loc}^{n}(t) \rangle  \simeq \frac{1}{(2v)^n}, ~~~~\text{as }\gamma t \to 0.
\label{alph-0-mom-eq-3-11}
\end{align}
Quite remarkably, we find that even when $t \to 0^+$, $\langle T_{loc}^{n}(t) \rangle$ has a non-vanishing value as shown in Eq. \eqref{alph-0-mom-eq-3-11}. Note that this is in contrast to the case of the Brownian particle where all moments vanish in this limit. On the other hand, in the limit $\gamma \to \infty$, $v \to \infty$ keeping $\mathfrak{D}_0=\frac{v^2}{2\gamma}$ fixed, we find that $\bar{Q}(p,s) \simeq \frac{2 \mathfrak{D}_0}{\sqrt{s}(p+2 \sqrt{s \mathfrak{D}_0})}$ substituting which in Eq. \eqref{alph-0-mom-eq-1} gives
\begin{align}
\langle T_{loc}^{n}(t) \rangle  \simeq \frac{n!}{\Gamma \left( \frac{n}{2}+1\right)} \left(\frac{t}{4 \mathfrak{D}_0} \right)^{\frac{n}{2}},~~\text{as }\gamma t,~v\to \infty.
\label{alph-0-mom-eq-3-22}
\end{align}
Expectedly, this represents the moments of $T_{loc}$ for the Brownian motion with diffusion constant $\mathfrak{D}_0$ \cite{SabhapanditS2006}.

\subsubsection{$P\left(T_{loc},t\right)$ for $\alpha =0$}
We now compute the probability distribution $P(T_{loc},t)$ for which we invert the Laplace transform $\bar{Q}(p,s)$ in Eq. \eqref{alph-0-eq-9}. The expression of $P(T_{loc},t)$ can be formally written as
\begin{align}
P\left(T_{loc},t\right) &= \mathcal{I}_{s \to t}^{-1}~~ \mathcal{I}_{p \to T_{loc}}^{-1} \left[\frac{(\lambda _s +s)e^{-\frac{p}{2 v}}}{s \left(\lambda _s +s +\gamma - \gamma e^{-\frac{p}{v}} \right)}\right], \nonumber \\
& = \mathcal{I}_{s \to t}^{-1}~~ \mathcal{I}_{p \to T_{loc}}^{-1} \left[\frac{\lambda _s-s}{s \gamma} \sum _{m=0}^{\infty} \left( \frac{\gamma}{\lambda _s +s+\gamma}\right)^{m} \right.\nonumber \\
& ~~~~~~~~~~~~~~~~~~~\times \Bigg. e^{-\frac{(2m+1)p}{2v}}\Bigg],
\label{alph-0-dist-eq-1}
\end{align}
where $\mathcal{I}_{s \to t}^{-1}~\left( \mathcal{I}_{p \to T_{loc}}^{-1}\right)$ is the inverse Laplace tranformation from $s \to t~(p \to T_{loc})$ and is defined in Eq. \eqref{inverse-lap}. Also, in going from first to second line, we have written the series expansion for the denominator. First we perform the inversion with respect to $p$ which is of the form $e^{-\beta p}$ with $\beta = \frac{2m+1}{2v}$ and which is equal to $\sim \delta \left(T_{loc}-\beta \right)$. Using this in Eq. \eqref{alph-0-dist-eq-1}, we get
\begin{align}
P\left(T_{loc},t\right) &= \mathcal{I}_{s \to t}^{-1} \left[ \frac{\lambda _s-s}{s \gamma} \sum _{m=0}^{\infty} \left( \frac{\gamma}{\lambda _s +s+\gamma}\right)^{m} \right. \nonumber \\
& ~~~~~~~~~~\times \left.\delta \left(T_{loc}-\frac{2m+1}{2v} \right) \right].
\label{alph-0-dist-eq-2}
\end{align}
We are now left with the inverse Laplace transformation with respect to $s$. To perform this inversion, we use the following identity \cite{bateman}:
\begin{align}
\mathcal{I}_{s \to t}^{-1} \left[ \frac{\lambda _s-s}{s \gamma}  \left( \frac{\gamma}{\lambda _s +s+\gamma}\right)^{m} \right]=e^{-\gamma t} \left[I_m(\gamma t) +I_{m+1}(\gamma t) \right].
\label{new-dist-0p0-eq-1}
\end{align}
Using this formula in Eq. \eqref{alph-0-dist-eq-2}, we get the expression of $P\left(T_{loc},t\right)$ as written in Eq. \eqref{alph-0-dist-eq-3}.
In Fig. \ref{prob-alph-0p0-pic-1} (left panel), we have compared the expression of $P\left(T_{loc},t\right)$ with the same obtained from the numerical simulations. Excellent match between them validates the expression in Eq. \eqref{alph-0-dist-eq-3}. To understand various $\delta$- function terms in Eq. \eqref{alph-0-dist-eq-3}, we look at the trajectories that give rise to these terms. In Fig. \ref{prob-alph-0p0-pic-1} (right panel), we have shown the trajectories that give rise to $\delta \left(T_{loc}-\frac{1}{2 v} \right)$ term (red) and $\delta \left(T_{loc}-\frac{3}{2 v} \right)$ term (blue). The $\delta \left(T_{loc}-\frac{1}{2 v} \right)$ term arises from those trajectories for which the RTP, starting from the origin, does not cross the origin till time $t$. The weight of this term will simply be the survival probability (from origin) whose expression is provided by $\mathcal{S}_0(t)$ in Eq. \eqref{alph-0-dist-eq-3333}. This expression of survival probability was also obtained in \cite{Malakar2018,Singh2020}. Similarly, the $\delta \left(T_{loc}-\frac{3}{2 v} \right)$ term in Eq. \eqref{alph-0-dist-eq-3} integrates the contribution of those trajectories for which particle crosses the origin once starting from the origin. The coefficient of this term $\mathcal{S}_1(t)$ is just the probability that the RTP crosses the origin once till time $t$. Extending the same argument, one gets the term $\mathcal{S}_m(t)~\delta \left(T_{loc}-\frac{2m+1}{2 v} \right)$ when the RTP crosses the origin $m$- times till time $t$. Interestingly, as a by-product, we have obtained $\mathcal{S}_m(t)$ which denotes the probability that the RTP crosses the origin $m$-times till time $t$ starting from the origin (see Appendix \ref{appen-der-smt} for derivation).\\
We now look at the expression of $P\left(T_{loc},t\right)$ in various limits of $\gamma t$. In the limit $\gamma t \to 0$, all $I_{m}(\gamma t \to 0)=0$ except $m=0$ for which $I_{0}(\gamma t \to 0)=1$. Hence in the R.H.S. of Eq. \eqref{alph-0-dist-eq-3}, only the $m=0$ term contributes and the expression of $P\left(T_{loc},t\right) $ becomes  
\begin{align}
P\left(T_{loc},t\right) \simeq \delta \left(T_{loc}-\frac{1}{2v} \right),~~~~~~\text{for } \gamma t \to 0.
\label{alph-0-dist-eq-4}
\end{align}
Using this expression, it is straightforward to verify that $\langle T^{n}_{loc}(t) \rangle \simeq \frac{1}{(2v)^n}$ which matches with the previously obtained result in Eq. \eqref{alph-0-mom-eq-3-11}. Let us now consider $P(T_{loc},t)$ in the limit $\gamma t \to \infty$. In this case the relevant scaling limit is $v T_{loc} \to \infty$ keeping $\frac{v T_{loc}}{\sqrt{\gamma t}}$ fixed to get a nontrivial expression. One can then replace the summation in the R.H.S. of Eq. \eqref{alph-0-dist-eq-3} by intergral performing which gives $P(T_{loc},t)\simeq v e^{-\gamma t} I_{v T_{loc}}(\gamma t)$. We next use the asymptotic form $I_{v T_{loc}}(\gamma t) \simeq \frac{e^{\gamma t}}{\sqrt{2 \pi \gamma t}} e^{-\frac{v^2 T_{loc}^2}{2 \gamma t}}$  and insert this in the expression of $P(T_{loc},t)$ to get
\begin{align}
P\left(T_{loc},t\right) \simeq \sqrt{\frac{4 \mathfrak{D}_0}{\pi t}} e^{-\frac{\mathfrak{D}_0 T_{loc}^2}{t}},~~~~~\text{as }\gamma t\to \infty, 
\label{alph-0-dist-eq-6}
\end{align} 
where $\mathfrak{D}_0 = \frac{v^2}{2 \gamma}$. Once again we emphasise that this expression is valid in the limits $\gamma t \to \infty$, $v T_{loc} \to \infty$ keeping $\frac{v T_{loc}}{\sqrt{\gamma t}}$ fixed. Since the particle behaves like a Brownian particle in these limits, Eq. \eqref{alph-0-dist-eq-6} represents the local time distribution for a Brownian particle in one dimension with diffusion constant $\mathfrak{D}_0$ \cite{SabhapanditS2006}.
\begin{figure*}[t]
  \centering
  \subfigure{\includegraphics[scale=0.3]{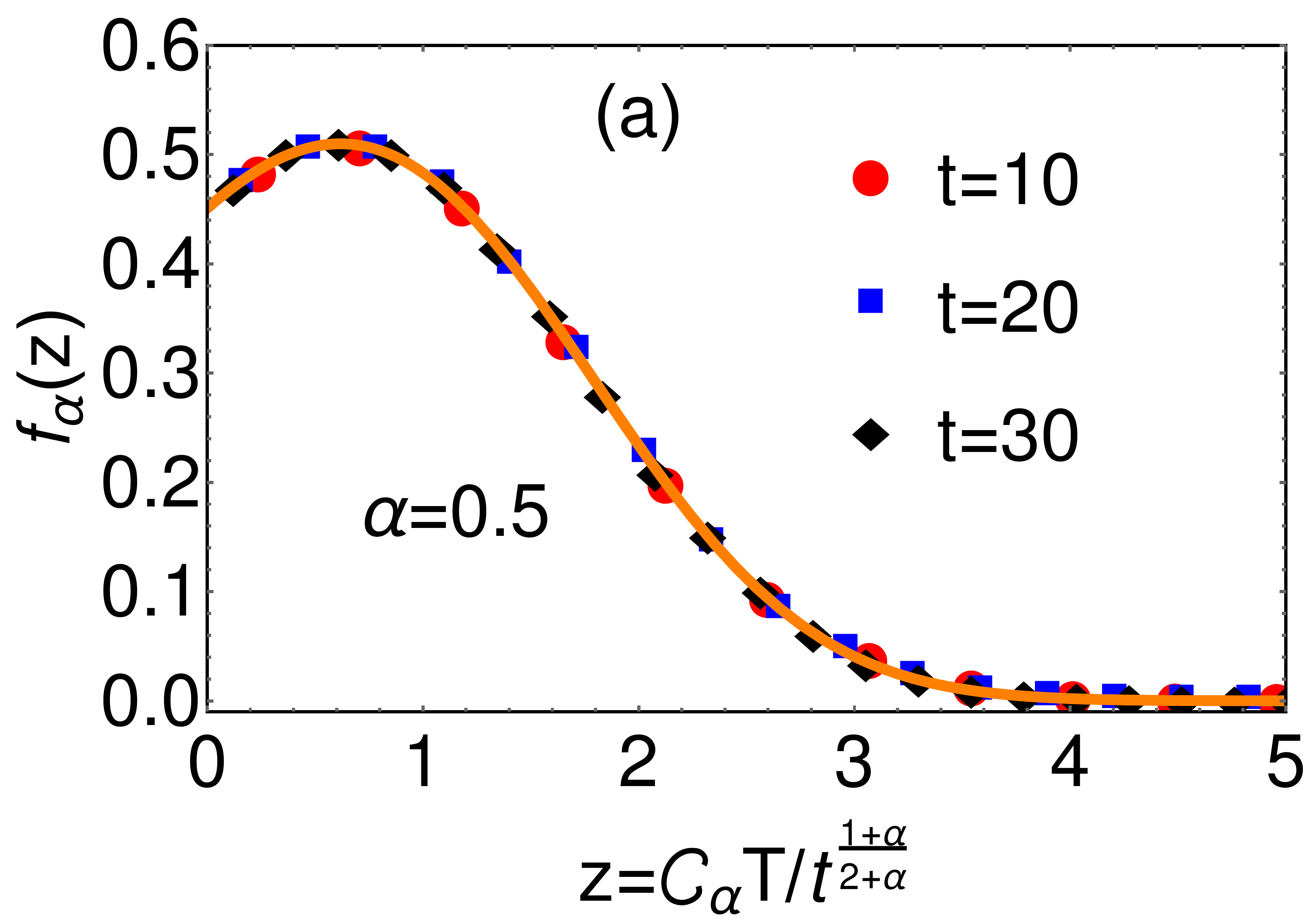}}
  \subfigure{\includegraphics[scale=0.3]{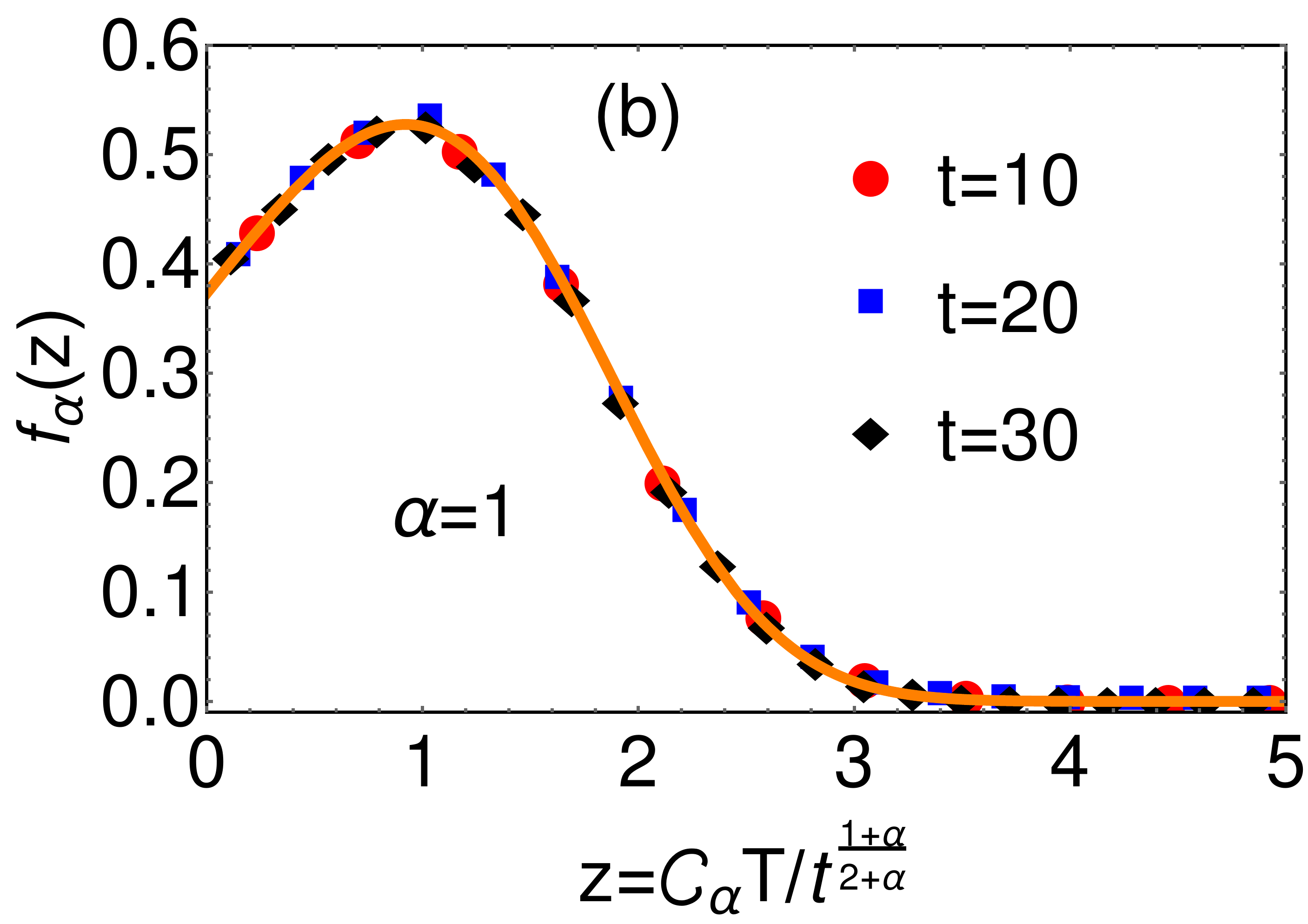}}
  \caption{(a) Comparision of the scaling function $f_{\alpha}(z)$ in Eq. \eqref{alph-neq-eq-13} (shown by solid line) with the same obtained via numerical simulations (shown by different symbols) for $\alpha =0.5$ and for different values of $t$. We have chosen $v=1,~\gamma=1$ and $l=1$. (b) Same is done for $\alpha =1$.}
\label{alp-neq-prob}  
\end{figure*}

\subsection{Case II: General $\alpha$}
\label{gen-alpha}
Let us now look at the statistics of $T_{loc}$ for the general $\alpha$ for which it is difficult to solve Eq.~\eqref{Q-eq-1} exactly for arbitrary $t$. On the other hand, the path-counting analysis used in the previous section to obtain the distribution of $T_{loc}$ for $\alpha =0$ in Eq. \eqref{alph-0-dist-eq-3} can also be extended to the general $\alpha$ case with a different form of $\mathcal{S}_m(t)$. However, getting $\mathcal{S}_m(t)$ for general $\alpha$, again turns out to be a challenging task although the Laplace transfrom  of it with respect to $t$ can be formally written in terms of the Laplace transform of the first passage probabilities as illustrated in Appendix \ref{appen-der-smt} (see Eq. \eqref{appen-der-smt-eq-3}).
In this case, we instead focus at  the large $t$ case for which we can make some analytical progress. Looking at the equation satisfied by $\bar{Z}$ in Eq. \eqref{bFP-eq-Lap-3}, we see that it contains a decay term of the form $2 R(x_0)\bar{Z}+q \mathbb{I}_{\epsilon} \left(  x_0 \right)$ while the equation for $\bar{Q}$ contains only $q \mathbb{I}_{\epsilon} \left(  x_0 \right)$ term as can be seen from Eq. \eqref{bFP-eq-Lap-2}. This implies that $\bar{Z}$ decays with time much faster than $\bar{Q}$. Hence for large $t$, we can neglect the $s \bar{Z}$ term and rewrite Eqs. \eqref{bFP-eq-Lap-2} and \eqref{bFP-eq-Lap-3} as 
\begin{align}
&s \bar{H}-1 = v \partial _{x_0} \bar{Z} - q \mathbb{I}_{\epsilon} \left(  x_0 \right) \bar{H}, \label{new-bFP-eq-Lap-2}\\
& v \partial _{x_0} \bar{H} -2 R(x_0)\bar{Z}- q \mathbb{I}_{\epsilon} \left(  x_0 \right) \bar{Z} \simeq 0.
\label{new-bFP-eq-Lap-3}
\end{align}
Since we are interested in computing the distribution of $T_{loc}$, we change $q \to \frac{p}{2 \epsilon}$ in these two equations. Now taking the limit $\epsilon \to 0$, one can replace  $\frac{p}{2\epsilon} \mathbb{I}_{\epsilon} \left(  x_0 \right) \to p \delta (x_0) $ and $\bar{H}\left( \frac{p}{2 \epsilon}, x_0,s\right) \to \bar{Q}(p,x_0,s)$ (as indicated in Eq. \eqref{new-infi-eq-6}) where $2\bar{Q}(p,x_0,s) =\bar{Q}_+(p,x_0,s)+\bar{Q}_-(p,x_0,s)$. Using this in Eqs. \eqref{new-bFP-eq-Lap-2} and \eqref{new-bFP-eq-Lap-3} yields
\begin{align}
&s \bar{Q}-1 = v \partial _{x_0} \bar{Z} - p \delta(x_0) \bar{Q}, \label{new-bFP-eq-Lap-4}\\
& v \partial _{x_0} \bar{Q} -2 R(x_0)\bar{Z}- p\delta (x_0)\bar{Z} \simeq 0, 
\label{new-bFP-eq-Lap-5} 
\end{align}
where $\bar{Z}(p,x_0,s)$ is to be interpreted as $2\bar{Z}(p,x_0,s) =\bar{Q}_+(p,x_0,s)-\bar{Q}_-(p,x_0,s)$. To solve Eqs. \eqref{new-bFP-eq-Lap-4} and \eqref{new-bFP-eq-Lap-5}, we note that if the particle initially starts from the origin, then the statistics of $T_{loc}$ is insensitive to the choice of the initial velocity direction and accordingly, we have $Q_+(p,x_0 =0,s) =Q_-(p,x_0 =0,s)$. This essentially asserts the condition that $\delta (x_0)\bar{Z}(p,x_0,s) =0$. With this result, we substitute $\bar{Z} \simeq \frac{v}{2 R(x_0)}\partial _{x_0} \bar{Q}$ from Eq. \eqref{new-bFP-eq-Lap-5} into Eq. \eqref{new-bFP-eq-Lap-4} and get a differential equation only for $\bar{Q}(p,x_0,s)$ as
\begin{align}
s \bar{Q}-1 \simeq  v^2 \frac{\partial}{\partial x_0} \left(\frac{1}{2 R(x_0)} \frac{\partial \bar{Q}}{\partial x_0}\right) - p \delta (x_0) \bar{Q}.
\label{alph-neq-eq-1}
\end{align}
We emphasise that the approximate equality indicates that this equation is valid only at large $t$. A similar approximation has been recently used in \cite{Singh2020} to study the position distribution and survival probability of RTP.

For $x_0 \neq 0$ we, from Eq.~\eqref{alph-neq-eq-1}, have
\begin{align}
s \bar{Q}-1 \simeq  \mathfrak{D}_{\alpha} \frac{\partial}{\partial x_0} \left(\frac{1}{|x_0|^{\alpha}} \frac{\partial \bar{Q}}{\partial x_0}\right),
\label{alph-neq-eq-2}
\end{align}
where $\mathfrak{D}_{\alpha} =\frac{v^2 l^{\alpha}}{2 \gamma}$. To turn this equation into a homogeneous form, we make the transformation
\begin{align}
\bar{Q}(p,x_0,s) = \frac{1}{s} + \mathcal{U}(p,x_0,s),
\label{alph-neq-eq-3}
\end{align}
and rewrite Eq. \eqref{alph-neq-eq-2} in terms of $\mathcal{U}$ as
\begin{align}
s ~\mathcal{U} \simeq  \mathfrak{D}_{\alpha} \frac{\partial}{\partial x_0} \left(\frac{1}{|x_0|^{\alpha}} \frac{\partial \mathcal{U}}{\partial x_0}\right).
\label{alph-neq-eq-4}
\end{align}
Recall that the boundary condition is $\bar{Q}(p,x_0 \to \pm \infty,s) = \frac{1}{s}$, which now gets translated to $\mathcal{U}(p,x_0 \to \pm \infty,s)=0$. The general solutions of Eq. \eqref{alph-neq-eq-4} are given in terms of the modified bessel functions of first kind and second kind as $|x_0|^{\frac{1+\alpha}{2}}I_{\frac{1+\alpha}{2+\alpha}} \left(\frac{2|x_0|^{\frac{2+\alpha}{2}}}{2+\alpha} \sqrt{\frac{s }{\mathfrak{D}_{\alpha}}} \right)$ and $|x_0|^{\frac{1+\alpha}{2}}K_{\frac{1+\alpha}{2+\alpha}} \left(\frac{2|x_0|^{\frac{2+\alpha}{2}}}{2+\alpha} \sqrt{\frac{s }{\mathfrak{D}_{\alpha}}} \right)$. However the first solution diverges for $|x_0| \to \infty$. Hence we are left only with the second solution as
\begin{align}
\mathcal{U}(p,x_0,s) \simeq \mathcal{B}(p,s) |x_0|^{\frac{1+\alpha}{2}}K_{\frac{1+\alpha}{2+\alpha}} \left(\frac{2|x_0|^{\frac{2+\alpha}{2}}}{2+\alpha} \sqrt{\frac{s }{\mathfrak{D}_{\alpha}}} \right),
\label{alph-neq-eq-5}
\end{align}
where $\mathcal{B}(p,s)$ is a $x_0$- independent function that remains to be computed. Translating this solution to get $\bar{Q}(p,x_0,s)$ from Eq. \eqref{alph-neq-eq-3}
\begin{align}
\bar{Q}(p,x_0,s) \simeq \frac{1}{s}+\mathcal{B}(p,s) |x_0|^{\frac{1+\alpha}{2}}K_{\frac{1+\alpha}{2+\alpha}} \left(\frac{2|x_0|^{\frac{2+\alpha}{2}}}{2+\alpha} \sqrt{\frac{s }{\mathfrak{D}_{\alpha}}} \right). 
\label{alph-neq-eq-6}
\end{align}
Next to compute $\mathcal{B}(p,s)$, we use Eq. \eqref{alph-neq-eq-1} and integrate it from $-\eta$ to $\eta$ and take $\eta \to 0$. This yields the discontinuity relation
\begin{align}
\left(\frac{1}{|x_0|^{\alpha}}\frac{ \partial \bar{Q}}{\partial x_0} \right) _{x_0= 0^+}-\left(\frac{1}{|x_0|^{\alpha}} \frac{\partial \bar{Q}}{\partial x_0} \right) _{x_0 = 0^-}  =\frac{ p}{\mathfrak{D}_{\alpha}} ~\bar{Q}(p, 0, s).
\label{alph-neq-eq-7}
\end{align}
Substituting the form of $\bar{Q}(p,x_0,s)$ from Eq. \eqref{alph-neq-eq-6} in this discontinuity relation, we get the solution of $\mathcal{B}(p,s)$ as
\begin{align}
\mathcal{B}(p,s) = \frac{1}{s} - \frac{\mathcal{C}_{\alpha}}{s^{\frac{1}{2+\alpha}}\left(p+\mathcal{C}_{\alpha} s^{\frac{1+\alpha}{2+\alpha}}\right)},
\label{alph-neq-eq-8hh}
\end{align}
where 
\begin{align}
\mathcal{C}_{\alpha} = 2 \left(\frac{\mathfrak{D}_{\alpha}}{(2+\alpha)^{\alpha}}\right)^{\frac{1}{2+\alpha}} \frac{\Gamma \left(\frac{1}{2+\alpha} \right)}{\Gamma \left(\frac{1+\alpha}{2+\alpha} \right)}.
\label{alph-neq-eq-911}
\end{align}
Finally inserting this form of $\mathcal{B}(p,s)$ in Eq. \eqref{alph-neq-eq-6} and setting $x_0=0$, we obtain
\begin{align}
\bar{Q}(p,s) \simeq \frac{\mathcal{C}_{\alpha}}{s^{\frac{1}{2+\alpha}}\left(p+\mathcal{C}_{\alpha} s^{\frac{1+\alpha}{2+\alpha}}\right)},
\label{alph-neq-eq-8}
\end{align}
where once again we have used the notation $\bar{Q}(p,s)$ for $\bar{Q}(p,0,s)$. We next use $\bar{Q}(p,s)$ to calculate the moments and distribution separately. 
\subsubsection{Moments $\langle T_{loc}^{n}(t) \rangle$ for general $\alpha$}
We first calculate the moments for general $\alpha$ case. We again emphasise that for $x_0=0$, the local time statistics is independent of the initial velocity direction which yields $\bar{Q}(p,s)=\bar{Q}_{\pm}(p,s)$. Following the definition of $\bar{Q}(p,s)$ in Eq. \eqref{LT-eq-1}, the moments can be written in terms of $\bar{Q}(p,s)$ as
\begin{align}
\langle T_{loc}^{n}(t) \rangle = \mathcal{I}_{s \to t}^{-1} \left[ (-1)^n \left( \frac{\partial ^n \bar{Q}(p,s)}{\partial p^n}\right)_{p=0}\right],
\label{alph-neq-eq-9}
\end{align}
where $\mathcal{I}_{s \to t}^{-1}$ is the inverse Laplace transformation. Inserting the expression of $\bar{Q}(p,s)$ from Eq. \eqref{alph-neq-eq-8}, we get
\begin{align}
\langle T_{loc}^{n}(t) \rangle &\simeq \mathcal{I}_{s \to t}^{-1} \left[ \frac{n!}{\mathcal{C}_{\alpha}^n  s^{\frac{1+(n+1)(1+\alpha)}{2+\alpha}}}\right].\label{alph-neq-eq-1023} 
\end{align}
Using the inverse Laplace transformation $\mathcal{I}_{s \to t}^{-1} [s^{-\nu}] =\frac{t^{\nu-1}}{\Gamma(\nu)}$ with $\nu > 0$ in Eq. \eqref{alph-neq-eq-1023}, we find the expression of moments as written in Eq. \eqref{alph-neq-eq-10}. It is worth mentioning that for general $\alpha$, the typical fluctuations of $T_{loc}$ scales with time as $T_{loc} \sim t^{\frac{1+\alpha}{2+\alpha}}$ for large values of $t$. This behaviour correctly reduces to the $\sqrt{t}$ scaling for $\alpha =0$ which corresponds to the homogeneous RTP model.

In Fig. \ref{alp-neq-mom}, we have compared our analytical result for moments with results obtained from the numerical simulations. We have plotted the first three moments for $\alpha = 0.5$. Although there is deviation between them for small $t$, the match becomes better as we go to higher values of $t$.
\subsubsection{Scaling form of $P(T_{loc},t)$ at large $t$ for general $\alpha$}
We next proceed to calculate $P(T_{loc},t)$ by inverting $\bar{Q}(p,s)$ in Eq. \eqref{alph-neq-eq-8} with respect to $p$ and $s$. Formally, the expression reads
\begin{align}
P(T_{loc},t) &\simeq \mathcal{I}_{s \to t}^{-1} \mathcal{I}_{p \to T_{loc}}^{-1}\left[ \frac{\mathcal{C}_{\alpha}}{s^{\frac{1}{2+\alpha}}\left(p+\mathcal{C}_{\alpha} s^{\frac{1+\alpha}{2+\alpha}}\right)}\right], \nonumber \\
& \simeq \mathcal{I}_{s \to t}^{-1} \left[ \frac{\mathcal{C}_{\alpha}}{s^{\frac{1}{2+\alpha}}} e^{-\mathcal{C}_{\alpha} T_{loc} s^{\frac{1+\alpha}{2+\alpha}}}\right].
\label{alph-neq-eq-11}
\end{align} 
In going from first line to second line, we have performed the Laplace inversion with respect to $p$. To obtain the distribution in the $t$ space, one needs to  perform the inversion with respect to $s$. For the purpose of maintaining  the continuity of our discussion, we have relegated the details of this inversion to Appendix \ref{inverse-Lap}. Using Eq.~\eqref{inverse-Lap-new-eq-5} and performing some simple manipulations we find that the distribution possesses the scaling form 
\begin{align}
P(T_{loc},t) \simeq \frac{\mathcal{C}_{\alpha}}{t^{\frac{1+\alpha}{2+\alpha}}} ~f_{\alpha} \left( \frac{\mathcal{C}_{\alpha} T_{loc}}{t^{\frac{1+\alpha}{2+\alpha}}}\right),
\label{alph-neq-eq-122}
\end{align}
where the scaling function $f_{\alpha}(z)$ is given in Eq. \eqref{alph-neq-eq-13}.
In Fig. \ref{alp-neq-mom}, we have illustrated this scaling behaviour and plotted the corresponding scaling function for two different values of $\alpha$ ($\alpha =0.5$ and $\alpha=1$). For each $\alpha$, we have obtained the simulation data for three different times and which collapse over the scaling function $f_{\alpha}(z)$ quite nicely.

Although the series summation in Eq. \eqref{alph-neq-eq-13} can be performed analytically for some values of $\alpha$, however it is difficult to do so for general $\alpha$. For example when $\alpha =0$, the summation can be carried out and we recover the result in Eq. \eqref{alph-0-dist-eq-6} for $P(T_{loc},t)$. On the other hand, when $\frac{\mathcal{C}_{\alpha} T_{loc}}{t^{\frac{1+\alpha}{2+\alpha}}}$ fixed, we can get a simplified expression of $f_{\alpha}(z)$ by using the saddle point approximation to evaluate the Bromwich integral in Eq. \eqref{alph-neq-eq-11}. We refer to Appendix \ref{saddle-f} for the detailed derivation of this approximate expression using saddle point method and present only the final result here as
\begin{align}
f_{\alpha}(z) \simeq \frac{z^{\frac{\alpha}{2}}}{\sqrt{2 \pi}} \frac{(1+\alpha)^{\frac{\alpha}{2}}}{(2+\alpha)^{\frac{\alpha-1}{2}}} ~ \text{exp}\left( -\frac{(1+\alpha)^{1+\alpha}}{(2+\alpha)^{2+\alpha}}~ z^{2+\alpha}\right).
\label{alph-neq-eq-14}
\end{align}
We remark that this expression for $f_{\alpha}(z)$ works only for large $z$ and fails for smaller values of $z$.

\begin{figure}[h]
\includegraphics[scale=0.3]{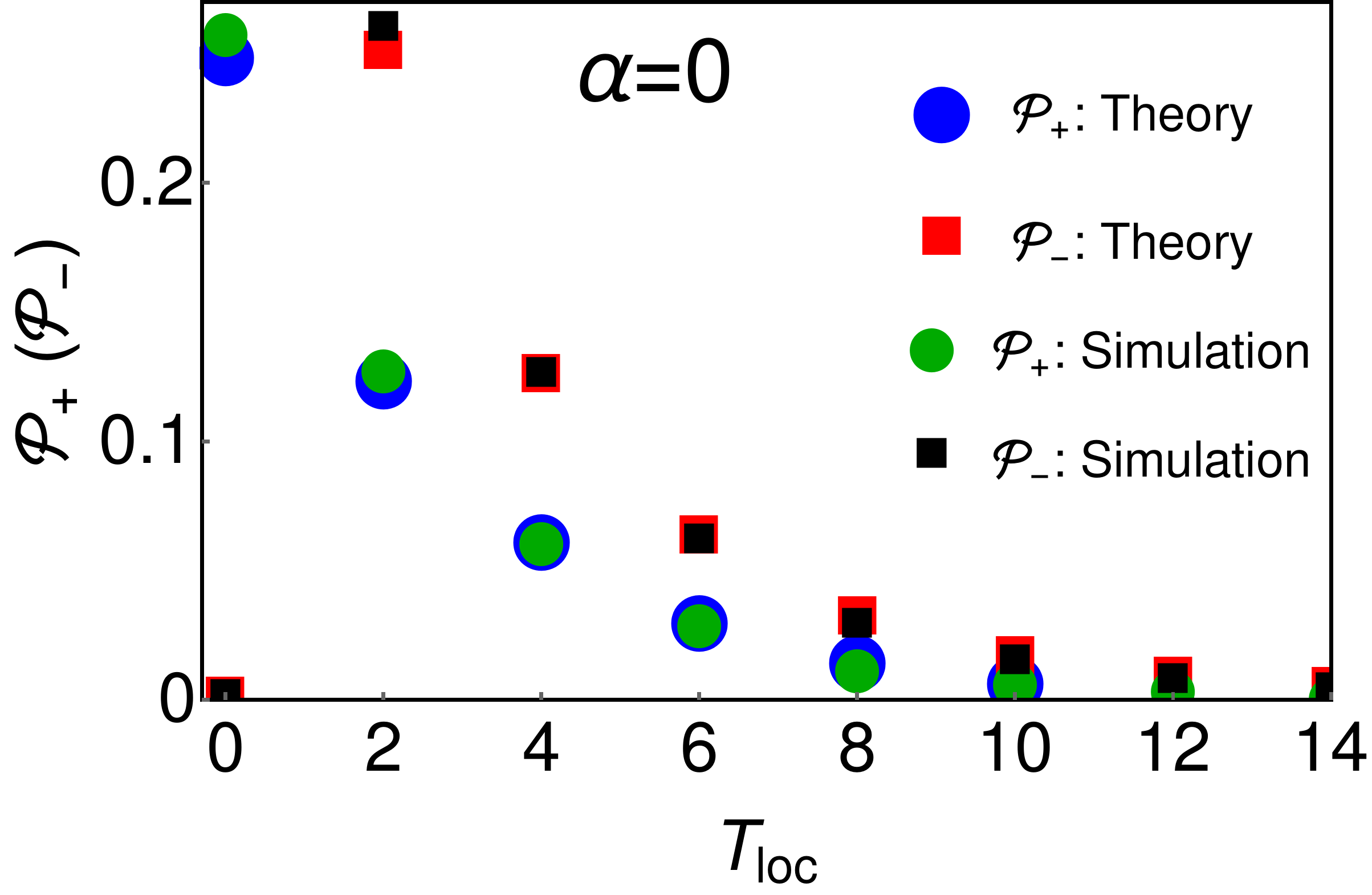}
\centering
\caption{Comparision of $\mathcal{P}_{\pm}(T_{loc})$ in Eqs. \eqref{abs-0-eq-10} and \eqref{abs-0-eq-11} for $\alpha = 0$ with that from the numerical simulations. We have taken $v=1,~\gamma=1$ and $M=1$. For simulation, we have taken $\epsilon = 0.001$}
\label{abs-alp-0p0}
\end{figure}

\section{LOCAL TIME IN PRESENCE OF AN ABSORBING WALL}
\label{loc-abs-wall}
In the previous section, we dealt with the statistics of $T_{loc}$ when the particle moves on an infinite line. However, in many physical settings, the time spent by a particle before getting absorbed becomes important. For example, in biological systems, the time spent by a molecule at some interior point of the cell before getting absorbed by the cell boundary may be of prime interest \cite{Redner2001}. Similarly, in chemical reactions where a catalytic agent interacts  with a reactant inside a bounded domain \cite{Pal2019}, one may be interested in the knowledge of the time spent by the agent with the reactant before it comes out of the domain as the yield of the product may depend on this time. Motivated from these applications, we study the statistical properties of the local time in presence of an absorbing wall.

Consider the RTP moving in one dimension with absorbing wall at $x= M$ $(M >0)$. Starting from $x_0 \leq M$, we define $T_{loc}$ as the time spent by the particle in the vicinity of $x=0$ before it hits the wall. This means that $T_{loc}$ is the time spend about the orgin before the first passage time $t_f$:
\begin{align}
T_{loc} = \int _{0}^{t_f} \delta \left( x(\tau)\right) d\tau.
\label{abs-eq-1}
\end{align}
In the context of Brownian motion, this quantity and similar path functionals till $t_f$ have been studied in detail. Such path functionals in the literature are called first passage functionals \cite{Majumdar2005}. 

Intuitively, the definition in Eq.~\eqref{abs-eq-1} can be understood as follows:
\begin{align}
&T_{loc} = \lim _{\epsilon \to 0^+} \frac{T_{2 \epsilon}}{2 \epsilon},~\text{with }\label{abs-eq-2}\\
& T_{2 \epsilon} = \int _{0}^{t_f} \mathbb{I} _{\epsilon} \left( x(\tau)\right)  d\tau.
\label{abs-eq-3}
\end{align}
Let us denote the distribution of $T_{loc}$ in this case as $\mathcal{P}_{\pm}(T_{loc}, x_0)$ which is different than $P_{\pm}(T_{loc}, x_0,t)$ written for the case of infinite line.
As done before, once again we define the Laplace transform of  $\mathcal{P}_{\pm}(T_{loc}, x_0)$ with respect to $T_{loc}$ as
\begin{align}
\mathcal{Q}_{\pm} (p,x_0) = \int _{0}^{\infty}d T_{loc}~ e^{-p T_{loc}}~\mathcal{P}_{\pm} (T_{loc},x_0).
\label{abs-eq-3p3}
\end{align}
In this case, we solve this problem by computing the distribution $\mathbb{P}_{\pm}(T_{2 \epsilon},x_0)$ of $T_{2 \epsilon}$ for a small finite $\epsilon$ and finally take $\epsilon \to 0$ appropriately as indicated in Eq. \eqref{abs-eq-2}. 
As before we define the Laplace transformation of $\mathbb{P}_{\pm}(T_{2 \epsilon},x_0)$ with respect to $T_{2 \epsilon}$ as 
\begin{align}
\mathbb{Q}_{\pm}(q,x_0)= \int _{0}^{\infty}d T_{2 \epsilon}~ e^{-q T_{2 \epsilon}}~\mathbb{P}_{\pm} (T_{2 \epsilon},x_0),
\label{abs-eq-3p3}
\end{align}
and write the corresponding backward Fokker-Planck equation as
\begin{align}
&v \partial _{x_0}\mathbb{Q}_+ - R(x_0) \mathbb{Q}_+ +R(x_0) \mathbb{Q}_--q \mathbb{I}_{\epsilon}(x_0) \mathbb{Q}_+=0, \label{abs-eq-3p6}\\
&v \partial _{x_0}\mathbb{Q}_-- R(x_0) \mathbb{Q}_+ +R(x_0) \mathbb{Q}_-+q \mathbb{I}_{\epsilon} (x_0)\mathbb{Q}_-=0. \label{abs-eq-3p7}
\end{align}
To solve these equations, we need to supplement them with appropriate boundary conditions. The boundary conditions are:
\begin{align}
&\mathbb{Q}_+(q, x_0 = M)=1, \label{abs-eq-3p8}\\
&\mathbb{Q}_{\pm}(q, x_0 \to -\infty) < \infty .
\label{abs-eq-3p99}
\end{align}
The first boundary condition arises due to the fact that if the particle starts initially from $M$ with $+v$ velocity, then it gets absorbed at the very next instant which means $T_{2 \epsilon}=0$, i.e. $\mathbb{P}(T_{2 \epsilon},M) = \delta \left(T_{2 \epsilon} \right)$ Translating this in terms of $\mathbb{Q}_+(q,x_0)$ via Eq. \eqref{abs-eq-3p3} gives the condition in Eq. \eqref{abs-eq-3p8}. To understand the second boundary condition, note that for $x_0 \to -\infty$, $t_f$ also diverges. However, the time spent inside the inteval $[-\epsilon, \epsilon]$ does not necessarily diverge or become zero. Hence we expect $\mathcal{P}_{\pm}(T_{2 \epsilon},x_0)$ to remain finite and non-zero even when $x_0 \to -\infty$, which in turn results in the boundary condition in Eq. \eqref{abs-eq-3p99}.

One can, in principle, solve these equations \eqref{abs-eq-3p6} and \eqref{abs-eq-3p7} along with the boundary conditions and obtain $\mathbb{Q}_{\pm}(q,x_0)$. Once this is established, it is straightforward to get the solution of $\mathcal{Q}_{\pm}(p,x_0)$ by inserting $T_{loc}$ from Eq. \eqref{abs-eq-2} in $\mathbb{Q}_{\pm}(q,x_0)$ and taking $\epsilon \to 0$ limit in Eq. \eqref{abs-eq-3p3}. The exact relation reads
\begin{align}
\mathcal{Q}_{\pm}(p,x_0) = \lim _{\epsilon \to 0} \mathbb{Q}_{\pm} \left(\frac{p}{2 \epsilon}, x_0 \right).
\label{abs-eq-3p9}
\end{align}
In the next section, we solve Eqs. \eqref{abs-eq-3p6} and \eqref{abs-eq-3p7} to get $\mathbb{Q}_{\pm}(q,x_0)$ and then use Eq. \eqref{abs-eq-3p9} to obtain $\mathcal{Q}_{\pm}(p,x_0)$. We consider first the $\alpha =0$ case which is followed by  $\alpha \neq 0$ case in the subsequent sections.

\subsection{Case I: $\alpha =0$}
\label{abs-alp}
We rewrite  the backward equations in \eqref{abs-eq-3p6} and \eqref{abs-eq-3p7} for $\alpha =0$ \emph{i.e.} $R(x_0)=\gamma$
\begin{align}
&~~~~v \partial _{x_0}\mathbb{Q}_+ -\gamma \mathbb{Q}_+ +\gamma \mathbb{Q}_--q \mathbb{I}_{\epsilon}(x_0) \mathbb{Q}_+=0, \label{abs-0-eq-1}\\
&-v \partial _{x_0}\mathbb{Q}_-+ \gamma \mathbb{Q}_+ -\gamma \mathbb{Q}_--q \mathbb{I}_{\epsilon} (x_0)\mathbb{Q}_-=0, \label{abs-0-eq-2}
\end{align}
where $\mathbb{I}_{\epsilon}(x_0)$ is $1$ if $-\epsilon < x_0 < \epsilon$ and $0$ otherwise. It is easy to solve these differential equations along with the boundary conditions in Eqs. \eqref{abs-eq-3p8} and \eqref{abs-eq-3p99}. The final solution reads 
\begin{widetext}
\begin{align}
\mathbb{Q}_+(q,x_0) = 
\begin{cases}
\mathbb{B}(q), ~~~~~~~~~~~~~~~~~~~~~~~~~~~~~~~~~~~\text{if }-\infty <x_0 <-\epsilon, \\
\mathbb{E}(q) e^{-\frac{\lambda(q)x}{v}}+\mathbb{F}(q) e^{\frac{\lambda(q)x}{v}}, ~~~~~~~~~\text{if }-\epsilon <x_0 <\epsilon,\\
\mathbb{C}(q)(x_0-M)+1,~~~~~~~~~~~~~~~~~~\text{if } \epsilon <x_0 <M.
\end{cases}
\label{abs-0-eq-5}.
\end{align}
\begin{align}
\mathbb{Q}_-( q, x_0) = 
\begin{cases}
\mathbb{B}(q), ~~~~~~~~~~~~~~~~~~~~~~~~~~~~~~~~~~~~~~~~~~~~~~~~~~~~~~~~~~~~\text{if }-\infty <x_0 <-\epsilon, \\
\mathbb{E}(q) \frac{\lambda(q)+\gamma +q}{\gamma}e^{-\frac{\lambda(q)x}{v}}+\mathbb{F}(q)  \frac{-\lambda(q)+\gamma +q}{\gamma}e^{\frac{\lambda(q)x}{v}}, ~~~~~~~~~\text{if }-\epsilon <x_0 <\epsilon,\\
\mathbb{C}(q)(x_0-M-\frac{v}{\gamma})+1,~~~~~~~~~~~~~~~~~~~~~~~~~~~~~~~~~~~~~\text{if } \epsilon <x_0 <M.
\end{cases}
\label{abs-0-eq-6}.
\end{align}
\end{widetext}
where $\lambda(q) = \sqrt{q(q+2 \gamma)}$ and $\mathbb{B}(q),~\mathbb{C}(q),~\mathbb{E}(q)$ and $\mathbb{F}(q)$ are independent of $x_0$ but depends on $q$. To evaluate these functions, we use the continuity of $\mathbb{Q}_{\pm}(q,x_0)$ at $x_0 = \pm \epsilon$ along with the boundary conditions at $x_0=M$ and at $x_0 \to -\infty$. These conditions give rise to four linear equations for $\mathbb{B}(q),~\mathbb{C}(q),~\mathbb{E}(q)$ and $\mathbb{F}(q)$ which can be solved to get these functions. Since we are interested in computing the distribution for $x_0 = \epsilon$, we here need only the expression of $\mathbb{C}(p)$. Also, since our goal is to compute the $\mathcal{Q}_{\pm}(p,x_0)$ using $\mathbb{Q}(q,x_0)$ via Eq. \eqref{abs-eq-3p9}, we provide here the expression of $\mathbb{C}\left(\frac{p}{2 \epsilon}\right)$ in the limit $\epsilon \to 0$. The expression reads
\begin{align}
 \lim _{\epsilon \to 0}\mathbb{C}\left(\frac{p}{2 \epsilon}\right) = \frac{\gamma \left( e^{\frac{2 p}{v}}-1\right)}{e^{\frac{2q}{v}}(v+\gamma M)-\gamma M}.
\label{abs-0-eq-7}
\end{align}
Inserting this expression in the last equations in the set of Eqs. \eqref{abs-0-eq-5} and \eqref{abs-0-eq-6}, we get the expressions of $\mathcal{Q}_+\left(\frac{p}{2 \epsilon} , x_0 = \epsilon \right) $ in the limit $\epsilon \to 0$ as
\begin{align}
&\mathcal{Q}_+(p)= \lim_{\epsilon \to 0}  \mathbb{Q}_+\left(\frac{p}{2 \epsilon}, \epsilon \right)  = \frac{v}{v+\gamma M-\gamma M e^{-\frac{2 p}{v}}}, \label{abs-0-eq-8}\\
& \mathcal{Q}_-(p)= \lim_{\epsilon \to 0} \mathbb{Q}_-\left(\frac{p}{2 \epsilon} ,\epsilon\right)  = \frac{v e^{-\frac{2 p}{v}}}{v+\gamma M-\gamma M e^{-\frac{2 p}{v}}},
\label{abs-0-eq-9}
\end{align}
where for simplicity, we have written $\mathcal{Q}_{\pm}(p, 0)$ simply as $\mathcal{Q}_{\pm}(p)$. To obtain the distribution, we have to perform the inverse Laplace transformation from $p$ to $T_{loc}$. To perform this inversion, we use $\mathcal{I}_{p \to T_{loc}}^{-1} \left[e^{-\tau p} \right] = \delta (T_{loc}-\tau)$ for $\tau >0$ and put it in Eqs. \eqref{abs-0-eq-8} and \eqref{abs-0-eq-9}. Finally, we get $\mathcal{P}_{\pm}(T_{loc})$ as written in Eqs. \eqref{abs-0-eq-10} and \eqref{abs-0-eq-11}.

\begin{figure*}[]
  \centering
  \subfigure{\includegraphics[scale=0.38]{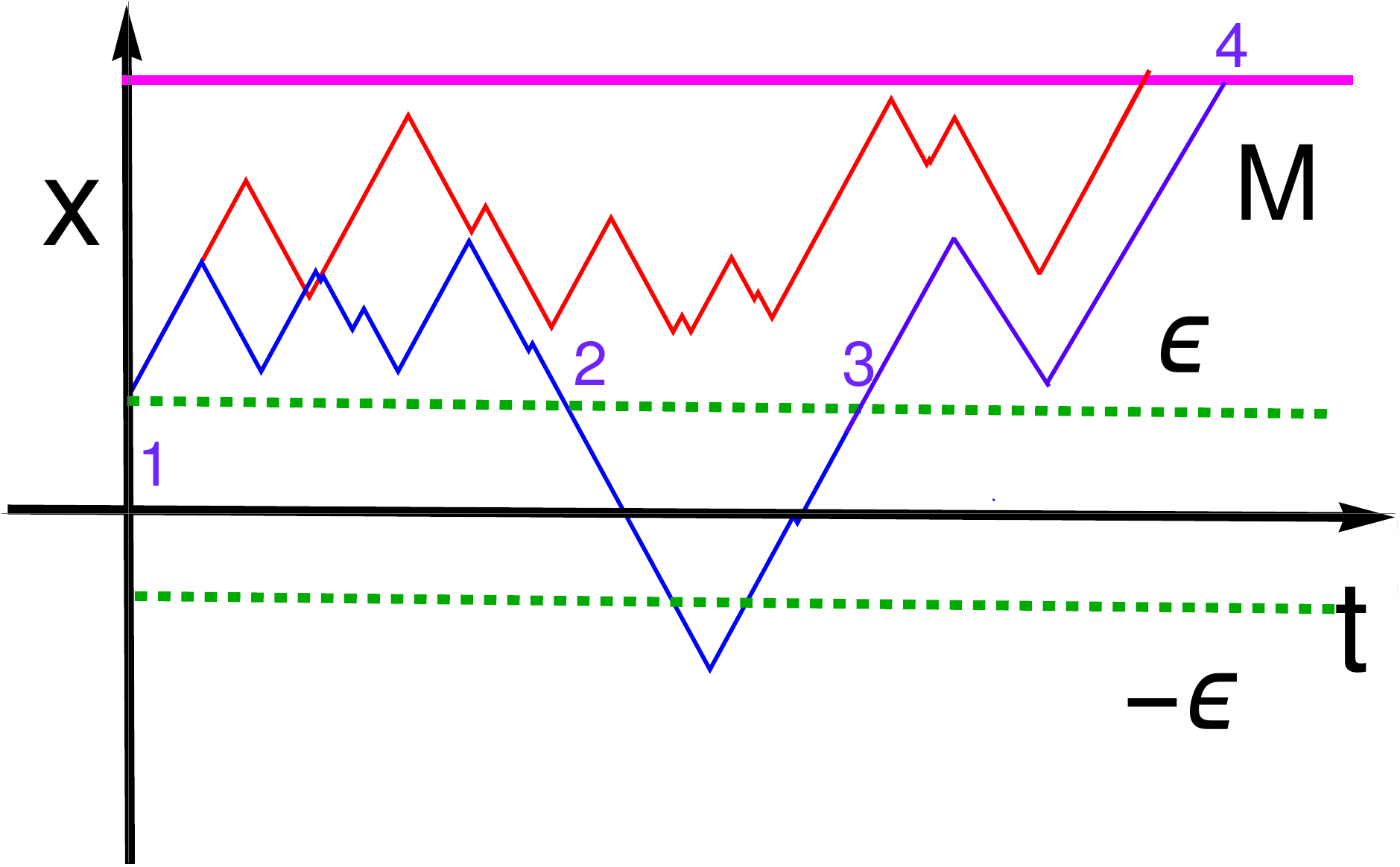}}
  \subfigure{\includegraphics[scale=0.35]{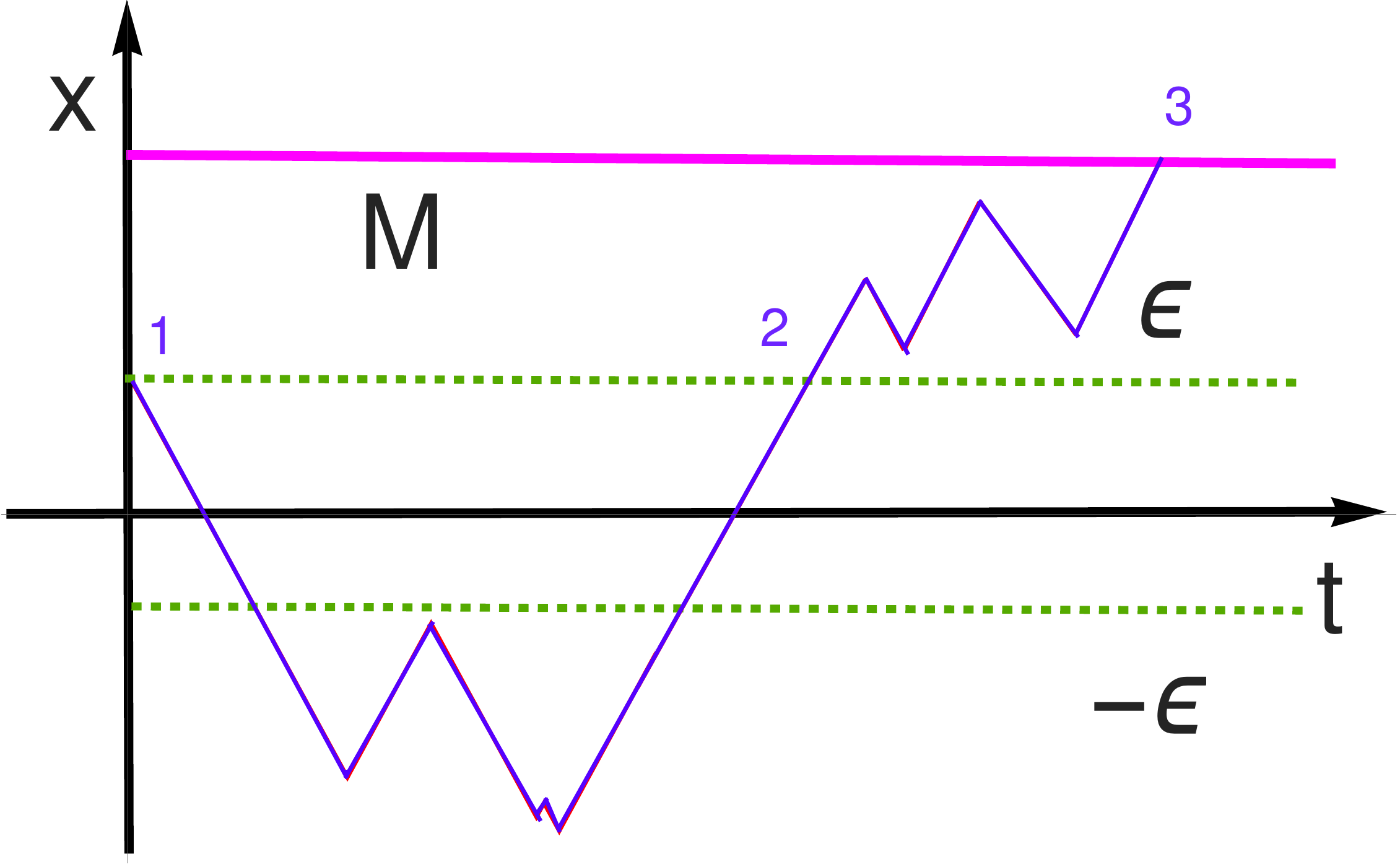}}
  \caption{Shematic of the realisations that give rise to different terms in $\mathcal{P}_{\pm}(T_{loc})$ in Eqs. \eqref{abs-0-eq-10} and \eqref{abs-0-eq-11} with the initial position $x_0 = \epsilon$. In both plots, the absorbing barrier at $x=M$ is shown in Magenta and the interval $[-\epsilon, \epsilon]$ is shown by dotted green lines. In \textit{left panel}, we have shown the typical realisations that give rise to $\delta (T_{loc})$ term (shown by red colour) and  $\delta \left(T_{loc}- \frac{2}{v}\right)$ term (shown by blue colour) when the RTP starts with $+v$ velocity. In \textit{right panel}, we have shown the same for the initial velocity $-v$. Note that $\delta (T_{loc})$ term does not arise for $\mathcal{P}_-(T_{loc})$ since the particle will definitely pass the interval $[-\epsilon, \epsilon]$ if it starts from $x_0 =\epsilon$ with $-v$ velcoity. }
\label{abs-scheme}  
\end{figure*}
In Figure \ref{abs-alp-0p0}, we have verfied the expressions of $\mathcal{P}_{\pm}(T_{loc})$ in Eqs. \eqref{abs-0-eq-10} and \eqref{abs-0-eq-11} against the numerical simulations. We observe excellent match for both cases. Note that while the summation for $\mathcal{P}_+(T_{loc})$ starts from $n=0$ and includes the term $\delta \left(T_{loc}-\frac{2}{v} \right)$, the summation for $\mathcal{P}_-(T_{loc})$ does not include this term. To understand this, let us look at the realisations that give rise to various $\delta$-function terms in $\mathcal{P}_{\pm}(T_{loc})$. The \textit{left panel} in Figure \ref{abs-scheme} shows two path realisations that give rise to $\delta\left(T_{loc} \right)$ term (red) and $\delta\left(T_{loc} -\frac{2}{v}\right)$ term (blue) starting from $x_0 = \epsilon$ with $+v$ velocity. The $\delta\left(T_{loc} \right)$ integrates the contributions of those realisations for which the particle reaches the absorbing wall $x=M$ without ever crossing the interval $[-\epsilon, \epsilon]$. The weight of this term will simply be the probability that the particle exits via $x=M$ without crossing the $x=0$. Denoting this exit probability by $\mathfrak{p}_0$, the expression of this probability was recently obtained in \cite{Malakar2018, Singh2020} as $\mathfrak{p}_0=\frac{v}{v+\gamma M}$ and accordingly we obtain $\mathfrak{p}_0~\delta\left(T_{loc} \right)$ as the contribution of such trajectories in the expression of $\mathcal{P}_+(T_{loc})$. Next we look at the  $\delta \left(T_{loc}-\frac{2}{v} \right)$ term in $\mathcal{P}_+(T_{loc})$ which originates from the blue colour path in the \textit{left panel} of Figure \ref{abs-scheme}. This term appears when the particle crosses the region $[-\epsilon, \epsilon]$ twice before getting absorbed and thereby contributing $T_{loc} = \frac{1}{2 \epsilon} \left(\frac{2 \epsilon}{v}+\frac{2 \epsilon}{v} \right) =\frac{2}{v}$. 

To compute the weight of such trajectories we decompose the blue trajectory (\textit{left panel}) as $[1 \to 2] \bigoplus [2 \to 3] \bigoplus [3 \to 4]$. Following the same argument as for $\delta (T_{loc})$, here also one gets the contribution of $1-\mathfrak{p}_0$ and $\mathfrak{p}_0$ from parts $1 \to 2$ and $3 \to 4$ respectively. On the other hand, the part $2 \to 3$ corresponds to the return probability for the RTP at $x=\epsilon$. Since the RTP motion is recurrent, this part will contribute unity which makes the overall weight as $\mathfrak{p}_0(1-\mathfrak{p}_0)\delta\left(T_{loc} -\frac{2}{v}\right)$. Substituting $\mathfrak{p}_0=\frac{v}{v+\gamma M}$, we get the coefficient of $\delta\left(T_{loc} -\frac{2}{v}\right)$ term in the expression of $\mathcal{P}_+(T_{loc})$ in Eq.~\eqref{abs-0-eq-10}.
Generalising this intuition to the case when the particle crosses the interval $[-\epsilon, \epsilon]$ $2 m$ times, one finds the corresponding contribution to $\mathcal{P}_+(T_{loc})$ is $\mathfrak{p}_0 (1-\mathfrak{p}_0)^{m} \delta\left(T_{loc} -\frac{2m}{v}\right)$. Note that the particle starting from $x_0=\epsilon$ has to cross the region $[-\epsilon,\epsilon]$ even number of times before it gets absorbed at $x=M>\epsilon$. As a result the $\delta$-functions in the expression of $\mathcal{P}_+(T_{loc})$ appear with $2m/v$ in the argument.

Next we try to understand the origin of various terms in $\mathcal{P}_-(T_{loc})$ in Eq. \eqref{abs-0-eq-11} from Figure \ref{abs-scheme} (\textit{right panel}). At first, recall that the particle will definitely enter the interval $[-\epsilon, \epsilon]$ and spend nonzero time there when it starts from $x_0 = \epsilon$ with $-v$ velocity and consequently one does not get $\delta (T_{loc})$ term. However, other terms still exist. For example, the Figure \ref{abs-scheme} (\textit{right panel}) shows a realisation which gives a contribution of $T_{loc} = \frac{2}{v}$ for this case. Once again, the path can be decomposed as $[1 \to 2] \bigoplus [2\to 3]$. While the path $1 \to 2$ gives unit contribution (since the motion is recurrent), the path $2 \to 3$ will contribute $\mathfrak{p}_0$ which makes the total contribution as $\mathfrak{p}_0~\delta\left(T_{loc} -\frac{2}{v}\right)$. Similarly if the particle crosses the interval $2m$ times, we get the overall contribution as $\mathfrak{p}_0 (1-\mathfrak{p}_0)^{m-1} \delta\left(T_{loc} -\frac{2m}{v}\right)$. Note that we have tacitly used the fact that in the limit $\epsilon \to 0$, the contribution of trajectories with tumbling inside the interval $x \in [-\epsilon, \epsilon]$ is zero.

Although in our analysis, we have chosen the initial position to be $\epsilon$, the overall physical argument can be extended to arbitrary initial position. For $x_0 = 0^-$ (\emph{i.e.} $x_0=\epsilon$ with $\epsilon \to 0^-$), it is possible to argue that the distribution of $T_{loc}$ is independent of the initial velocity and is given by 
\begin{align}
\mathcal{P}_\pm(T_{loc})=\sum_{m=0}^\infty \mathfrak{p}_0(1-\mathfrak{p}_0)^{m} \delta\left( T_{loc}-\frac{2m+1}{v}\right).
\label{ak-eq-1}
\end{align}
On the other hand for $x_0=0$ the distribution depends on the initial velocity and is given by 
\begin{align}
\begin{split}
\mathcal{P}_+(T_{loc})=\sum_{m=0}^\infty \mathfrak{p}_0(1-\mathfrak{p}_0)^{m} \delta\left( T_{loc}-\frac{4m+1}{2v}\right), \\
\mathcal{P}_-(T_{loc})=\sum_{m=0}^\infty \mathfrak{p}_0(1-\mathfrak{p}_0)^{m} \delta\left( T_{loc}-\frac{4m+3}{2v}\right). \\
\end{split}
\label{ak-eq-2}
\end{align}
Numerical verifications of these expressions are given in appendix \ref{appen-fig}

The path-analysis method seems more useful compared to solving backward Fokker Planck equation because the latter for general $x_0$ could be quite non-trivial. Also, this physical insight guides us to generalise the results of $\mathcal{P}_{\pm}(T_{loc})$ in Eqs. \eqref{abs-0-eq-10} and \eqref{abs-0-eq-11} to the other values of $\alpha$. On the other hand, solving the backward Fokker Planck equations for general $\alpha$ could be a challenging task. In the next section, we generalise the results of $\mathcal{P}_{\pm}(T_{loc})$ for $\alpha =0$ to the arbitrary values of $\alpha$.
\begin{figure*}[]
  \centering
  \subfigure{\includegraphics[scale=0.3]{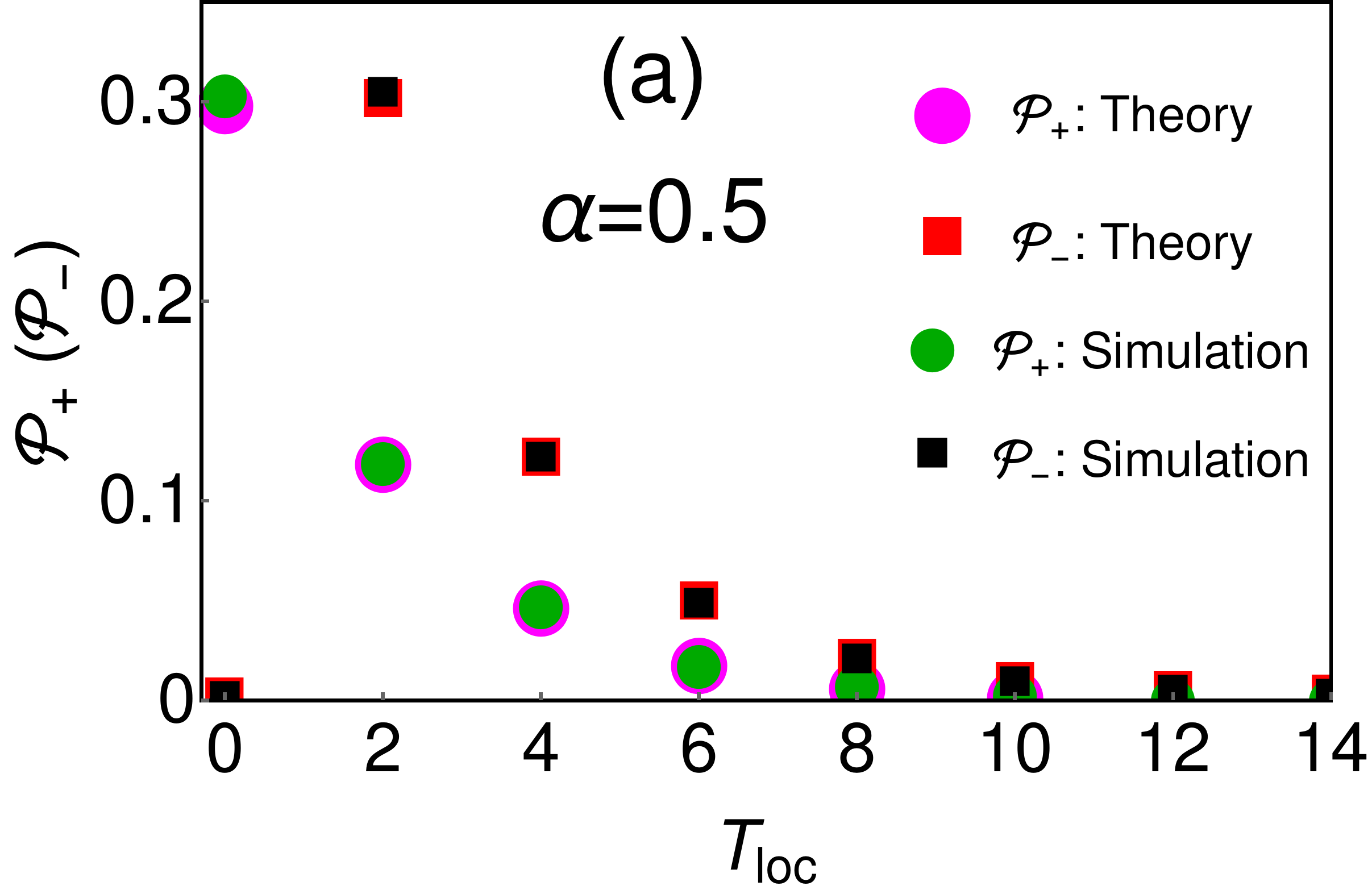}}
  \subfigure{\includegraphics[scale=0.3]{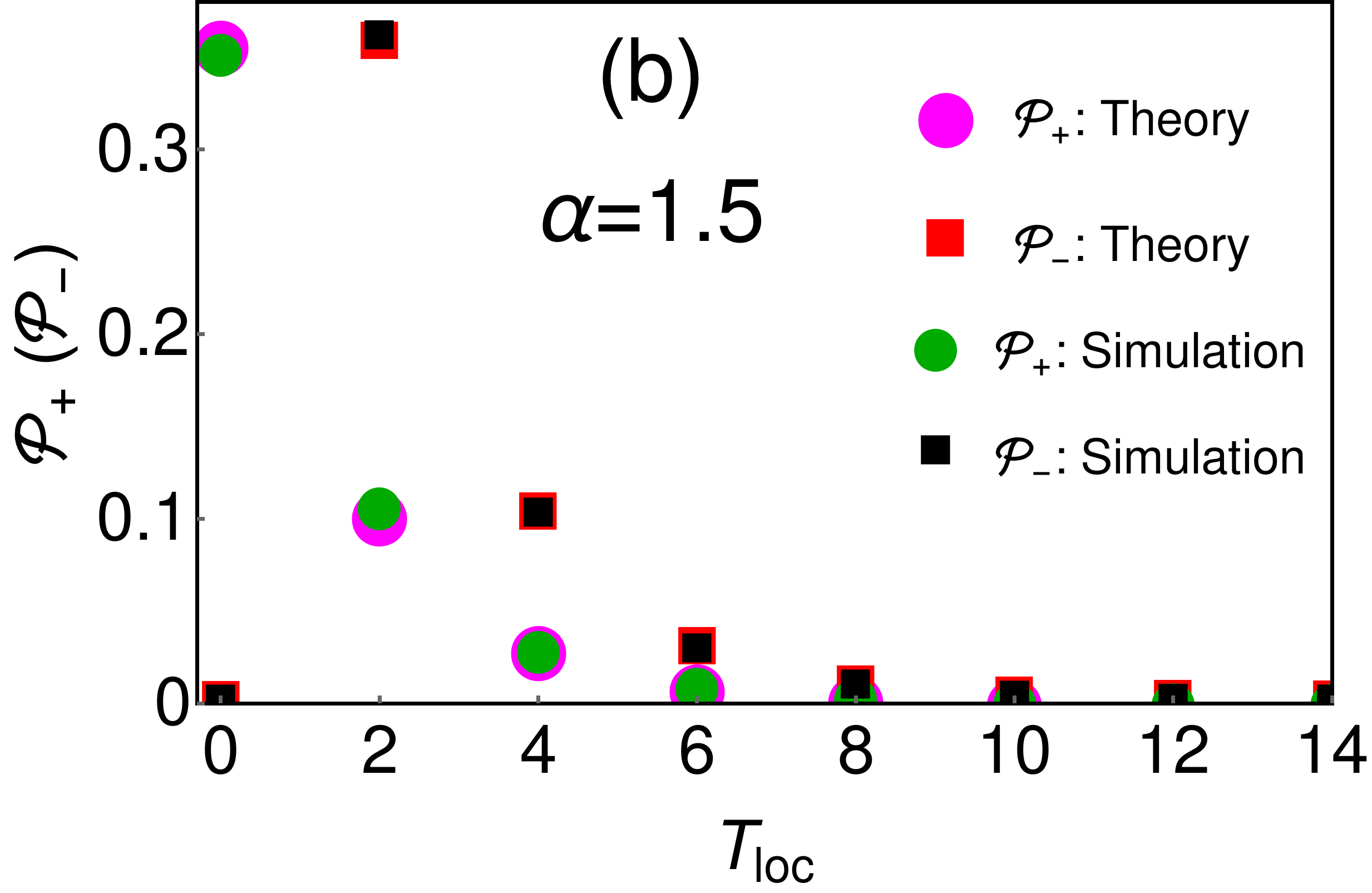}}
  \caption{Comparision of $\mathcal{P}_{\pm}(T_{loc})$ in Eqs. \eqref{abs-n0-eq-1} and  \eqref{abs-n0-eq-2} with the numerical simulations for two values of $\alpha$- (a) $\alpha = 0.5$ and (b) $\alpha=1.5$. For both plots, we have chosen $v=1,~\gamma=1,~M=1,~l=1$. For simulation, we have taken $\epsilon = 0.001$.}
\label{abs-non-zero-alpha}  
\end{figure*}
\subsection{Case II: General $\alpha$}
\label{abs-alp-gen}
We now look at the distribution of $T_{loc}$ for the general $\alpha$ in presence of an absorbing wall at $x=M$. 
Solving the backward Fokker Plank equations \eqref{abs-eq-3p6} and \eqref{abs-eq-3p7} analytically for arbitrary values of $\alpha~ (\geq 0)$ is highly difficult except for some special values of $\alpha$. 
In this section we show that the derivation of $\mathcal{P}_{\pm}(T_{loc})$ based on the trajectory analysis in Eqs. \eqref{abs-0-eq-10} and \eqref{abs-0-eq-11}  for $\alpha=0$ can be easily extended to the general values of $\alpha$. Following the physical arguments, for this case also, we can write the distributions $\mathcal{P}_{\pm}(T_{loc})$ as
\begin{align}
&\mathcal{P}_+(T_{loc})= \sum _{m=0}^{\infty} \mathfrak{p}_{\alpha} \left( 1-\mathfrak{p}_{\alpha} \right)^{m} \delta \left( T_{loc}-\frac{2 m}{v}\right), \label{abs-n0-eq-1}\\
&\mathcal{P}_-(T_{loc})= \sum _{m=1}^{\infty} \mathfrak{p}_{\alpha} \left( 1-\mathfrak{p}_{\alpha} \right)^{m-1} \delta \left( T_{loc}-\frac{2 m}{v}\right),
\label{abs-n0-eq-2}
\end{align} 
where we have chosen the initial position to be $x_0=\epsilon $ for simplicity. Also, $\mathfrak{p}_{\alpha }$ stands for the probability that the RTP starting from the origin exits from $x=M$ without touching the origin. The  expression of $\mathfrak{p}_{\alpha}$ for arbitrary $\alpha$ was obtained in \cite{Singh2020} which reads
\begin{align}
\mathfrak{p}_{\alpha} = \frac{v (1+\alpha)l^{\alpha}}{\gamma M^{1+\alpha}+v (1+\alpha)l^{\alpha}}.
\label{abs-n0-eq-3}
\end{align}
For $\alpha=0$, the expressions in Eqs. \eqref{abs-n0-eq-1} and \eqref{abs-n0-eq-2} reduce to that in Eqs. \eqref{abs-0-eq-10} and \eqref{abs-0-eq-11}. In Figure \ref{abs-non-zero-alpha}, we have compared these expressions with that from the numerical simulations for two non-zero values of $\alpha$. We observe excellent match of our analytical results with the numerical data.

\section{CONCLUSION}
\label{conclusion}
In this paper, we have studied the local time properties of an one dimensional run and tumble particle in inhomogeneous medium. The inhomogeneity was administered by taking the position-dependent rate $R(x)$ given in Eq. \eqref{rate-eq}. For $\alpha = 0$ (which corresponds to the homogeneous case), we have derived the moments generating function from which we computed exactly the first three moments of $T_{loc}$. Next, we derived the distribution of $T_{loc}$ for this case which consists of a series of appropriately weighted $\delta$ -functions at $\delta\left(T_{loc} - \frac{2m+1}{2v}\right)$, where $m=0,1,2,...$(see Eq. \eqref{alph-0-dist-eq-3}). While for $t >>\frac{1}{\gamma}$ and $v T_{loc} \to \infty$ keeping $\frac{v T_{loc}}{\sqrt{\gamma t}}$ fixed, this expression correctly converges to that of the one dimensional Brownian motion. However, in the opposite limit  $t <<\frac{1}{\gamma}$ the statistics is remarkably different from Brownian motion. For the latter scenario, we have shown that the distribution is $P(T_{loc},t) \simeq \delta\left(T_{loc}-\frac{1}{2 v} \right)$. Due to this, the moments have non-zero value as $t \to 0^+$ which is contrary to that of the Brownian motion where all moments vanish as $t \to 0^+$. The appearance of $\delta\left(T_{loc} - \frac{2m+1}{2v}\right)$ in  $P(T_{loc},t)$ can be understood from the path-conuting analyisis which also gives the weight of these $\delta$-functions. Based on this analysis, we showed that the weight of $\delta\left(T_{loc} - \frac{2m+1}{2v}\right)$ in  $P(T_{loc},t)$ is given in terms of the probability $\mathcal{S}_m(t)$ that the RTP visits the origin $m$-times till time $t$ starting from the origin. The exact expression of $\mathcal{S}_m(t)$ is given in Eq. \eqref{alph-0-dist-eq-3333}.

For general $\alpha >0$, performing an exact calculation turns out to be difficult. However for large $t$, we showed that $T_{loc}$ scales with respect to time as $T_{loc} \sim t^{\frac{1+\alpha}{2+\alpha}}$ and the distribution possesses a scaling behaviour of the form $P(T_{loc},t) \simeq \frac{\mathcal{C}_{\alpha}}{t^{\frac{1+\alpha}{2+\alpha}}} ~f_{\alpha} \left( \frac{\mathcal{C}_{\alpha} T_{loc}}{t^{\frac{1+\alpha}{2+\alpha}}}\right)$, with the scaling function $f_{\alpha}(z)$ given exactly in Eq. \eqref{alph-neq-eq-13}. We also computed all the moments of $T_{loc}$ for large $t$ and general $\alpha >0$.

The second part of our paper dealt with the local time spent by the particle near the origin in presence of an absorbing wall at $x=M~(>0)$. Starting from $x_0 = \epsilon$, we define $T_{loc}$ as the time (density) spent by the particle in the vicinity of the origin before it hits the wall. For $\alpha = 0$, we derived the exact probability distribution of $T_{loc}$ by solving the backward Fokker-Planck equations. As shown in Eqs. \eqref{abs-0-eq-10} and \eqref{abs-0-eq-11}, the distribution in this case also  consists of a series of $\delta$-functions at the $T_{loc}=\frac{2n}{v}$ where $n=0,1,2,3,...$ for $\mathcal{P}_+(T_{loc})$ and $n=1,2,3,...$ for $\mathcal{P}_-(T_{loc})$. In this case also, the coefficient of  $\delta\left(T_{loc} - \frac{2n}{v}\right)$ can be interpreted as the probability of $m$-th visits to the origin by the RTP starting from the origin till it gets absorbed. Quite remarkably, we found that this physical insight can also be extended to the case of the general $\alpha~(>0)$ which guided us to write exactly the distribution without solving the backward Fokker-Planck equations. These distributions are given in Eqs. \eqref{abs-n0-eq-1} and \eqref{abs-n0-eq-2} and verified numerically in fig.~\ref{abs-non-zero-alpha}.

In this paper, we have focused on one particular class of functionals called local time. Studying the statistical properties of other functionals is an interesting future direction. Recently in \cite{MajumdarMeerson2020}, the statistics of first passage functional of type $\int _0 ^{t_f} d \tau x^k (\tau)$ with $k>-2$ was studied for one dimensional Brownian motion with a drift where the distribution was shown to exhibit dynamical phase transition for some values of $k$. It would be interesting to see how these results get modified for RTP dynamics. Finally, in this work, we have focused only on the typcial fluctuations of $T_{loc}$ which was shown to scale at large time as $T_{loc} \sim t^{\frac{1+\alpha}{2+\alpha}}$. It remains a promising future direction to explore the atypical fluctuations of $T_{loc}$ and see if the corresponding distribution admits large deviation forms. Another interesting direction to extend the current work would be to study  the distribution of local  time for active Brownian particle.

\begin{acknowledgements}
PS acknowledges fruitful discussions with Soummyadip Basak. AK and PS acknowledge support of the Department of Atomic Energy, Government of India, under project no.12-R\&D-TFR-5.10-1100. AK acknowledges support from DST, Government of India grant under project No. ECR/2017/000634.
\end{acknowledgements}
\appendix
\section{Backward master equation for $Q_{\pm}(p,x_0,t)$}
 \label{BME-Q}
Here we provide a derivation of the Backward master equations in \eqref{new-infi-eq-5} for a general functional $Y\left[ x\left( t\right)\right]$ 
\begin{align}
 Y\left[ x\left( t\right)\right]= \int^t_0 d\tau U\left[ x\left( \tau\right)\right], \label{res1}
\end{align}
of the trajectory $\{x(\tau);~0 \leq \tau \leq t \}$ of  the RTP starting from $x_0 $ at $t =0$. Denoting the probability distribution of $Y$ as $G_{\sigma}(Y,x_0,t)$ where $\sigma = \pm 1$ is the direction of the initial velocity, we consider the characteristic function
\begin{align}
  H_{\sigma}(q,x_0,t)=\int^{\infty}_0 dY e^{-q Y} G_{\sigma}\left(Y,x_0,t\right) =\left< e^{-q Y}\right>_{\left(x_0,\sigma\right)},
\label{res2}
\end{align}
 where $<...>_{\left(x_0,\sigma\right)}$ denotes the average with initial position $x_0$ and velocity orientation $\sigma$. Following the definition in Eq.(\ref{res2}), we write for a small time interval $\Delta t$
\begin{align}
\begin{split}
  H_{\sigma}&\left(q,x_0,t+ \Delta  t \right)=\left< e^{-q \int^{t+\Delta t}_0 d\tau U\left[ x\left( \tau\right)\right]}\right>_{\left(x_0,\sigma\right)}\\
  &=\left< e^{-q \int^{\Delta t}_0 d\tau U\left[ x\left( \tau\right)\right]}\indent e^{-q \int^{t+\Delta t}_{\Delta t} d\tau U\left[ x\left( \tau\right)\right]} \right>_{\left(x_0,\sigma\right)}\\
  &=\left(1-q U(x_0)\Delta t\right)\left<e^{-q \int^{t+\Delta t}_{\Delta t} d\tau U\left[ x\left( \tau\right)\right]} \right>_{\left(x_0,\sigma\right)}.
\label{res3}
\end{split}
\end{align}
where $(x_0, \sigma)$ is the initial configuration.  In time interval $\Delta t$ the state $(x_0, \sigma)$ of the RTP can change  to $(x_0+\sigma v \Delta t, \sigma)$ with probability $1-R(x_0) \Delta t$ and to $(x_0, -\sigma)$ with probability $R(x_0) \Delta t$. We can write Eq.(\ref{res3}) as following:

\begin{widetext}
\begin{align}
 H_{\sigma}(q,x_0,t+\Delta t ) &=\left(1-q U\Delta t\right)\left[\left(1-R \Delta t\right)\langle e^{-q \int^{t+\Delta t}_{\Delta t} d\tau U\left[ x\left( \tau\right)\right]} \rangle_{\left(x_0+\sigma v \Delta t, \sigma\right)} +R \Delta t\left<e^{-q \int^{t+\Delta t}_{\Delta t} d\tau U\left[ x\left( \tau\right)\right]} \right>_{\left(x_0-\sigma v \Delta t, -\sigma\right)}\right], \nonumber \\ 
  &=\left(1-q U(x_0)\Delta t\right)\left[ \left(1-R(x_0) \Delta t\right) H_{\sigma}(q,x_0+\sigma v \Delta t,t )+R(x_0) \Delta t H_{-\sigma}(q,x_0,t )\right].
\end{align}
\end{widetext}
Taking the $\Delta t \to 0$ limit we obtain 
\begin{align}
\partial_t H_{\sigma}=\sigma v \partial_{x_0} H_{\sigma}- R(x_0) H_{\sigma}+ R(x_0) H_{-\sigma}-q U(x_0)H_{\sigma}.
\label{res5}
\end{align}
If we choose $U(x)= \mathbb{I}_{\epsilon}(x)$ then $Y$ represents the time spent in the interval $x \in [-\epsilon, \epsilon]$  and for this choice one obtains the differential equations \eqref{new-infi-eq-5}. 

\section{Derivation of $\bar{H}_{\pm}(q,x_0,s)$ in Eqs. \eqref{new-alp-0p0-eq-2} and \eqref{new-alp-0p0-eq-3}}
\label{sol-alph-0p0}
In this appendix, we solve the differential equation Eq. \eqref{alph-0-Q-eq-1} to  obtain the solutions of $\bar{H}_{\pm}(q,x_0,s)$ in Eqs. \eqref{new-alp-0p0-eq-2} and \eqref{new-alp-0p0-eq-3}. To begin with, we rewrite Eq. \eqref{alph-0-Q-eq-1} here as
\begin{align}
s \bar{H}-1 =  \frac{\partial}{\partial x_0} \left(\frac{v^2}{s+2 \gamma+q \mathbb{I}_{\epsilon}} \frac{\partial \bar{H}}{\partial x_0}\right) - q \mathbb{I}_{\epsilon} \left(  x_0 \right) \bar{H},
\label{appen-sol-alhpa-0p0-eq-1}
\end{align}
where $\mathbb{I}_{\epsilon}(x_0)$ is equal to $1$ for $-\epsilon <x_0 <\epsilon$ and $0$ otherwise. Let us first look at Eq. \eqref{appen-sol-alhpa-0p0-eq-1} when $-\infty <x_0<-\epsilon$ for which $\mathbb{I}_{\epsilon}(x_0)=0$. This differential equation then becomes
\begin{align}
s \bar{H}-1 =  \frac{v^2}{s+2 \gamma}\frac{\partial^2 \bar{H}}{\partial x_0^2}.
\label{appen-sol-alhpa-0p0-eq-2}
\end{align}
We now make the following transformation
\begin{align}
\bar{H}(q,x_0,s) = \frac{1}{s} + \mathcal{W}(q,x_0,s),
\label{appen-sol-alhpa-0p0-eq-3}
\end{align}
in Eq. \eqref{appen-sol-alhpa-0p0-eq-2} to obtain
\begin{align}
\frac{\partial^2 \mathcal{W}}{\partial x_0^2} = \frac{\lambda _s^2}{v^2}\mathcal{W},
\label{appen-sol-alhpa-0p0-eq-4}
\end{align}
where $\lambda _s = \sqrt{s(2 \gamma +s)}$. This is a simple equation which needs to be solved with appropriate boundary conditions. From Eqs. \eqref{barQ-s-bc} and \eqref{Q-Z-eq-1}, it is easy to see that $\bar{H}(q, x_0 \to \pm \infty,s)=\frac{1}{s}$ which, through Eq. \eqref{appen-sol-alhpa-0p0-eq-3} yields $\mathcal{W}(q, x_0 \to \pm \infty,s) =0$. It is now easy to solve Eq. \eqref{appen-sol-alhpa-0p0-eq-4} along with these boundary conditions to get
\begin{align}
\mathcal{W}(q,x_0,s) = \mathcal{B}_1(q,s) e^{\frac{\lambda_s}{v} x_0},
\label{appen-sol-alhpa-0p0-eq-5}
\end{align}
where $\mathcal{B}_1(q,s)$ is a $x_0$ - independent function that needs to be computed. Using this in Eq. \eqref{appen-sol-alhpa-0p0-eq-3} gives
\begin{align}
\bar{H}(q,x_0,s) = \frac{1}{s}+\mathcal{B}_1(q,s) e^{\frac{\lambda_s}{v} x_0}.
\label{appen-sol-alhpa-0p0-eq-6}
\end{align}
To compute $\bar{H}_{\pm}(q,x_0,s)$, we also need $\bar{Z}(q,x_0,s)$ as is evident from Eq. \eqref{Q-Z-eq-2}. To this aim, we use Eq. \eqref{bFP-eq-Lap-3} appropriately for $\alpha =0$ to write $\bar{Z}(q,x_0,s)$ in terms of $\bar{H}(q,x_0,s)$ as $\bar{Z}(q,x_0,s) = \frac{v}{s+2 \gamma} \partial _{x_0} \bar{H}$. Inserting the solution of  $\bar{H}(q,x_0,s)$ in Eq. \eqref{appen-sol-alhpa-0p0-eq-6} and using Eqs. \eqref{Q-Z-eq-1} and \eqref{Q-Z-eq-2}, we find
\begin{align}
&\bar{H}_+(q,x_0,s) = \frac{1}{s} + \mathcal{A}_1(q,s) e^{\frac{\lambda _s x_0}{v}}, \label{appen-sol-alhpa-0p0-eq-7}\\
&\bar{H}_-(q,x_0,s) = \frac{1}{s} + \mathcal{A}_1(q,s) \left( \frac{-\lambda _s +s+\gamma}{\gamma}\right)e^{\frac{\lambda _s x_0}{v}}.\label{appen-sol-alhpa-0p0-eq-8}
\end{align}
We emphasise that these two solutions are valid for $-\infty <x_0 <-\epsilon$ which are also written in Eqs. \eqref{new-alp-0p0-eq-2} and \eqref{new-alp-0p0-eq-3}. Also, we have defined a new function $\mathcal{A}_1(q,s)$ which depends on $\mathcal{B}_1(q,s)$. We next look at the solution of Eq. \eqref{appen-sol-alhpa-0p0-eq-1} when $\epsilon < x_0 <\infty$. Although it is possible to solve this equation explicitly, we, however, use the symmetry $\bar{H}_{\pm}(q,x_0,s) = \bar{H}_{\mp}(q,-x_0,s)$ to write the solutions in the region $\epsilon < x_0 <\infty$. Using Eqs. \eqref{appen-sol-alhpa-0p0-eq-7} and \eqref{appen-sol-alhpa-0p0-eq-8}, the solutions can be written as
\begin{align}
&\bar{H}_+(q,x_0,s) = \frac{1}{s} + \mathcal{A}_1(q,s)\left( \frac{-\lambda _s +s+\gamma}{\gamma}\right) e^{-\frac{\lambda _s x_0}{v}}, \label{appen-sol-alhpa-0p0-eq-9}\\
&\bar{H}_-(q,x_0,s) = \frac{1}{s} + \mathcal{A}_1(q,s)e^{-\frac{\lambda _s x_0}{v}}.\label{appen-sol-alhpa-0p0-eq-10}
\end{align}
These solutions are written in Eqs. \eqref{new-alp-0p0-eq-2} and \eqref{new-alp-0p0-eq-3} for $\epsilon < x_0 <\infty$. We now solve Eq. \eqref{appen-sol-alhpa-0p0-eq-1} for $-\epsilon <x_0 <\epsilon$ for which $\mathbb{I}_{\epsilon}(x_0)=1$. Proceeding as before,it is straightforward to get the solution as
\begin{align}
\bar{H}(q,x_0,s) = \frac{1}{s+q} + \mathcal{B}_3(q,s) e^{\frac{\lambda _q x_0}{v}}+ \mathcal{B}_4(q,s) e^{-\frac{\lambda _q x_0}{v}},
\label{appen-sol-alhpa-0p0-eq-11}
\end{align}
where $\lambda _q = \sqrt{(s+q)(s+q+2 \gamma)}$ and $\mathcal{B}_3(q,s)$ and $\mathcal{B}_4(q,s)$ are $x_0$- independent functions. Using Eq. \eqref{bFP-eq-Lap-3}, we find $\bar{Z}(q,x_0,s) = \frac{v}{s+2 \gamma+q} \partial _{x_0} \bar{H}$ which, through Eqs. \eqref{Q-Z-eq-1} and \eqref{Q-Z-eq-2} gives
\begin{align}
&\bar{H}_+(q,x_0,s) = \frac{1}{s+q} + \mathcal{A}_2(q,s) e^{\frac{\lambda _q x_0}{v}}+ \mathcal{A}_3(q,s) e^{-\frac{\lambda _q x_0}{v}}, \label{appen-sol-alhpa-0p0-eq-12} \\
& \bar{H}_-(q,x_0,s) =\frac{1}{s+q} + \mathcal{A}_2(q,s) \left(\frac{-\lambda _q+s+q+\gamma}{\gamma} \right)e^{\frac{\lambda _q x_0}{v}} \nonumber \\
 &~~~~~~~~~~~~~~~~~~~~~~+\left(\frac{\lambda _q+s+q+\gamma}{\gamma} \right)\mathcal{A}_3(q,s) e^{-\frac{\lambda _q x_0}{v}},
\label{appen-sol-alhpa-0p0-eq-13}
\end{align}
where we have defined two new functions $\mathcal{A}_2(q,s)$ and $\mathcal{A}_3(q,s)$ that depend on $\mathcal{B}_3(q,s)$ and $\mathcal{B}_4(q,s)$. Finally, we use the symmetry $\bar{H}_{\pm}(q,x_0,s) = \bar{H}_{\mp}(q,-x_0,s)$ again to write $\mathcal{A}_3(q,s)=\left(\frac{-\lambda _q+s+q+\gamma}{\gamma} \right)\mathcal{A}_2(q,s)$ and insert this in Eqs. \eqref{appen-sol-alhpa-0p0-eq-12} and \eqref{appen-sol-alhpa-0p0-eq-13} to get
\begin{align}
\bar{H}_{\pm}(q,x_0,s) = &\frac{1}{s+q} + \mathcal{A}_2(q,s) e^{\pm \frac{\lambda _q x_0}{v}} \nonumber\\
&+ \left(\frac{-\lambda _q+s+q+\gamma}{\gamma} \right)\mathcal{A}_2(q,s) e^{\mp \frac{\lambda _q x_0}{v}}.
\label{appen-sol-alhpa-0p0-eq-14}
\end{align}
This result is quoted in Eqs. \eqref{new-alp-0p0-eq-2} and \eqref{new-alp-0p0-eq-3} for $x_0 \in [-\epsilon, \epsilon]$.

\section{Derivation of $\mathcal{S}_m(t)$}
\label{appen-der-smt}
This appendix deals with the derivation of $\mathcal{S}_{m}(t)$ in Eq. \eqref{alph-0-dist-eq-3333}. Based on the path-counting analysis, we saw in Eq. \eqref{alph-0-dist-eq-3} that the distribution of $T_{loc}$ is given as 
\begin{align}
&P\left(T_{loc},t\right) = \sum _{m=0}^{\infty} \mathcal{S}_{m}(t)~ \delta \left(T_{loc}-\frac{2m+1}{2v} \right), \label{appen-der-smt-eq-1}
\end{align}
where $\mathcal{S}_{m}(t)$ was interpreted as the probability that the RTP returns to the origin $m$-times in time $t$ starting from the origin. In this appendix, we rigorously derive the expression of $\mathcal{S}_{m}(t)$ as given in Eq. \eqref{alph-0-dist-eq-3333} for $\alpha =0$. Let us assume that the RTP initially has $+$ velocity direction. The probability  $\mathcal{S}_{m}(t)$ can be formally written as
\begin{align}
\mathcal{S}_{m}(t)& = \int _{0}^{t} d \tau_1 F_+(\tau _1) \int_{0}^{t -\tau _1} d \tau _2 F_-(\tau _1)... \nonumber\\
&\times \int _{0}^{t-\sum _{i=0}^{m-1}\tau _i} d \tau _m F_{(-1)^{m-1}}(\tau _m)  \mathbb{S}_{(-1)^{m-1}} \left(t-\sum _{i=0}^{m}\tau _i\right),
\label{appen-der-smt-eq-2}
\end{align}
where $F_{\sigma}(\tau _i)$ is the probability distribution for the RTP to arrive at the origin, starting from origin with initial velocity direction $\sigma$, with $\sigma \in [+,-]$ and $\mathbb{S}_{\sigma}(t_i)$ is the probability that the RTP has not returned to the origin till time $t_i$ starting from the origin. Note that both these quantities are insensitive to the initial velocity direction which implies that the end result of $\mathcal{S}_m(t)$ will be independent of the initial velocity direction. To understand Eq. \eqref{appen-der-smt-eq-2}, we procede as follows. In time $[\tau _1, \tau _1 + d \tau _1]$, the RTP visits the origin for the first time starting from the origin with speed $v$ which contributes $F_+(\tau _1) d \tau _1$ in Eq. \eqref{appen-der-smt-eq-2}. In the further time $[\tau _1+\tau _2 , \tau_1+\tau _2 +d \tau _2]$, starting from the origin, the RTP visits the orgin for the second time. Since the particle was at the origin at time $\tau _1$ with $-v$ speed, we can interprete the second visit at time $\tau _1 +\tau _2$ as the first-arrival in the time $\tau _2$. with initial speed $-v$. This gives rise to $F_-(\tau _2) d \tau _2$ term in Eq. \eqref{appen-der-smt-eq-2}. Similarly, the $m$-th visit to the origin at time $\tau_1 +\tau_2 +\tau_3 +....+\tau _m$ can be interpreted as the first visit at time $\tau _m$ since at time $\tau_1 +\tau_2 +\tau_3 +....+\tau _{m-1}$, the particle was at the origin with speed $(-1)^{m-1} v$. Accordingly, we have $F_{(-1)^{m-1}}(\tau _m) d \tau _m$ term in Eq. \eqref{appen-der-smt-eq-2}. Once the particle has visited the origin $m$-times, the particle will not cross the origin in the remaining time $t-\tau_1 -\tau_2-...-\tau _m$ which gives rise to the survival probability term $\mathbb{S}_{(-1)^{m-1}} \left(t-\sum _{i=0}^{m}\tau _i\right)$ in Eq. \eqref{appen-der-smt-eq-2}. Finally, we have appropriately integrated over $\tau _i$ for all $i=1,2,3,...,m$.

Although Eq. \eqref{appen-der-smt-eq-2} looks a bit complicated, it possesses the convolution structure which can be used to simplify expression in terms of the Laplace variable $s$ with respect to $t$. Using the convolution theorem for Laplace transforms, the Laplace transform of $\mathcal{S}_{m}(t)$ (which we denote by $\bar{\mathcal{S}}_{m}(s)$) will just be the product of Laplace transforms of $F_{\sigma}(\tau)$'s and $\mathbb{S}_{\sigma}(\tau)$. The expression of  $\bar{\mathcal{S}}_{m}(s)$ reads
\begin{align}
\bar{\mathcal{S}}_{m}(s) = \left[ \bar{F}_{\sigma}(s)\right]^m \bar{\mathbb{S}}_{\sigma}(s),
\label{appen-der-smt-eq-3}
\end{align}
where $\bar{F}_{\sigma}(s)$ and $\bar{\mathbb{S}}_{\sigma}(s)$ are the Laplace transforms of $F_{\sigma}(\tau)$'s and $\mathbb{S}_{\sigma}(\tau)$ respectively. Since $F_{\sigma}(\tau)$'s and $\mathbb{S}_{\sigma}(\tau)$ are independent of $\sigma,$ the  we have dropped the dependence of initial velocity direction in Eq. \eqref{appen-der-smt-eq-3} and replaced them with $\sigma$. Coming back to Eq. \eqref{appen-der-smt-eq-3}, note that in deriving $\bar{\mathcal{S}}_{m}(s)$, we have not made any reference to $\alpha$ which means that Eqs. \eqref{appen-der-smt-eq-1},  \eqref{appen-der-smt-eq-2} and  \eqref{appen-der-smt-eq-3} remain true for all $\alpha ~(\geq 0)$. However, the exact form of $\bar{F}_{\sigma}(s)$ and $\bar{\mathbb{S}}_{\sigma}(s)$ is known only for $\alpha =0$ which read \cite{Malakar2018, Singh2020},
\begin{align}
& \bar{\mathbb{S}}_{\sigma}(s) = \frac{\lambda(s)-s}{s \gamma}, \label{appen-der-smt-eq-4}\\
& \bar{F}_{\sigma}(s)= \frac{\gamma}{s+\gamma + \lambda(s)},
\label{appen-der-smt-eq-5}
\end{align}
where $\lambda (s) = \sqrt{s(s+2 \gamma)}$. Inserting these expressions of $\bar{F}_{\sigma}(s)$ and $\bar{\mathbb{S}}_{\sigma}(s)$ in Eq. \eqref{appen-der-smt-eq-3}, we get
\begin{align}
\bar{\mathcal{S}}_{m}(s)  = \left( \frac{\lambda(s)-s}{s \gamma} \right) \left( \frac{\gamma}{s+\gamma + \lambda(s)}\right)^m.
\label{appen-der-smt-eq-6}
\end{align}
To perform the inverse Laplace transformation, we use Eq. \eqref{new-dist-0p0-eq-1} to get
\begin{align}
\mathcal{S}_{m}(t) =e^{-\gamma t}\left[I_m(\gamma t) +I_{m+1}(\gamma t) \right],
\label{appen-der-smt-eq-7}
\end{align}
which matches with the $\mathcal{S}_m(t)$ in Eq. \eqref{alph-0-dist-eq-3333}.

\section{Inverse Laplace transformation of $\frac{1}{s^{\frac{1}{2+\alpha}}}e^{-\mathcal{C}_{\alpha} T_{loc} s^{\frac{1+\alpha}{2+\alpha}}}$}
\label{inverse-Lap}
Here we will derive the inverse Laplace transformation of Eq. \eqref{alph-neq-eq-11} that leads to the scaling structure of $P(T_{loc},t)$ in Eq. \eqref{alph-neq-eq-122} for general $\alpha$. To this aim, we consider the following inverse Laplace transformation:
\begin{align}
\Lambda(\beta ,t,w)=\mathcal{I}_{s \to t}^{-1} \left( e^{-w s^{\beta}}\right).
\label{inverse-Lap-new-eq-1}
\end{align} 
Following \cite{Montroll1984}, this inverse Laplace transformation is given by
\begin{align}
\Lambda(\beta ,t,w)= -\frac{1}{\pi} \sum _{n=1}^{\infty} \frac{(-w)^n}{n!} \frac{\Gamma\left( 1+n \beta\right)}{t^{n \beta +1}} \sin \left(\pi n \beta \right).
\label{inverse-Lap-new-eq-2}
\end{align}
We next define another inverse Laplace transform as
\begin{align}
\mathcal{G}(\beta ,t,w) = \mathcal{I}_{s \to t}^{-1} \left( \frac{e^{-w s^{\beta}}}{s^{1-\beta}}\right).
\label{inverse-Lap-new-eq-3}
\end{align}
using the definition of inverse Laplace transform in Eq. \eqref{inverse-lap}, it is easy to show that
\begin{align}
\frac{\partial }{\partial t} \mathcal{G}(\beta ,t,w)&=\frac{\partial}{\partial w}\Lambda(\beta ,t,w), \\
& = \frac{1}{\pi}\sum _{n=1}^{\infty} \frac{n(-w)^{n-1}}{n!} \frac{\Gamma\left( 1+n \beta\right)}{t^{n \beta +1}} \sin \left(\pi n \beta \right),
\label{inverse-Lap-new-eq-4}
\end{align}
where we have used Eq. \eqref{inverse-Lap-new-eq-2} in going from first line to second line. Finally integrating Eq. \eqref{inverse-Lap-new-eq-4} from $t$ to $\infty$, we get
\begin{align}
\mathcal{G}(\beta ,t,w)=\frac{1}{\pi}\sum _{n=1}^{\infty} \frac{(-w)^{n-1}}{n! } \frac{\Gamma\left( 1+n \beta\right)}{\beta~t^{n \beta }} \sin \left(\pi n \beta \right).
\label{inverse-Lap-new-eq-5}
\end{align}
Substituting $\beta = \frac{1+\alpha}{2+\alpha}$ and $w = \mathcal{C}_{\alpha} T_{loc}$, we see that Eq. \eqref{inverse-Lap-new-eq-3} becomes the inverse Laplace transfrom of $\frac{1}{s^{\frac{1}{2+\alpha}}} e^{-\mathcal{C}_{\alpha} T_{loc} s^{\frac{1+\alpha}{2+\alpha}}}$. Using the form of $\mathcal{G}\left(\frac{1+\alpha}{2+\alpha} ,t,\mathcal{C}_{\alpha} T_{loc} \right)$ in Eq. \eqref{alph-neq-eq-11}, we obtain the scaling form of $P(T_{loc},t)$ in Eq. \eqref{alph-neq-eq-122}.
\begin{figure*}[t]
  \centering
  \subfigure{\includegraphics[scale=0.3]{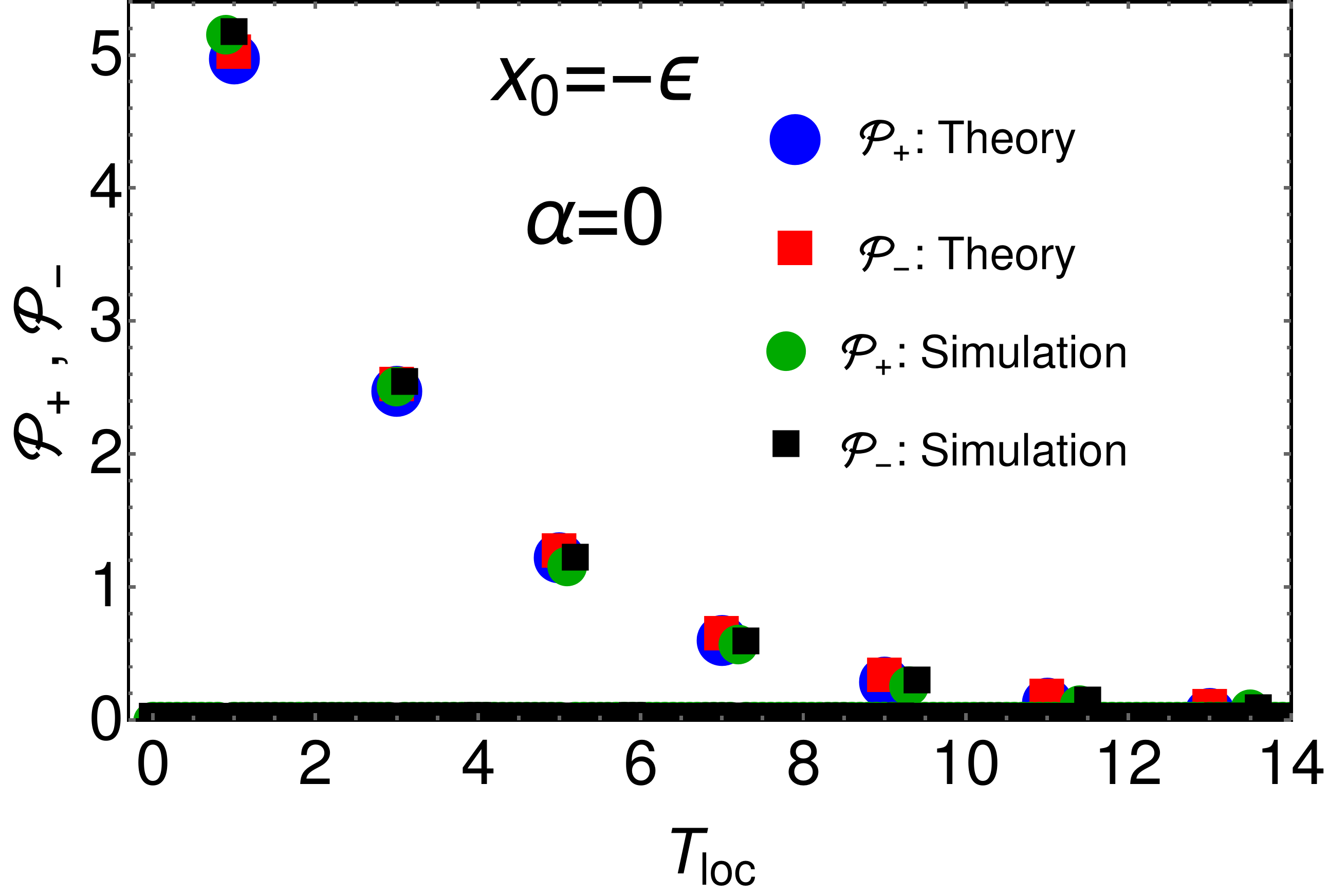}}
  \subfigure{\includegraphics[scale=0.3]{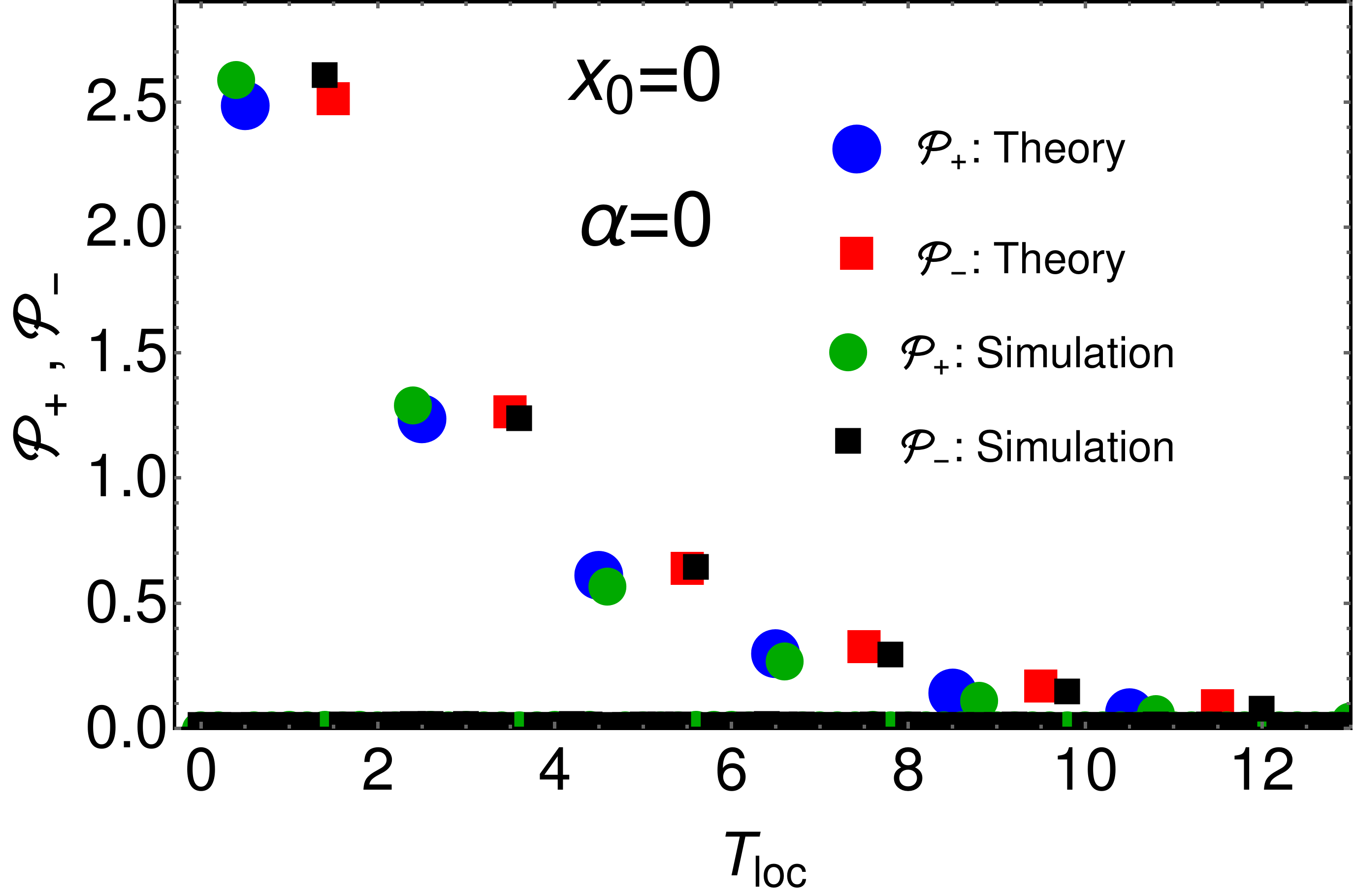}}
  \caption{Comparision of $\mathcal{P}_{\pm}(T_{loc})$ in Eqs. \eqref{ak-eq-1} (left panel) and  \eqref{ak-eq-2} (right panel) with the numerical simulations for two values of $\alpha$- (a) $\alpha = 0.5$ and (b) $\alpha=1.5$. For both plots, we have chosen $v=1,~\gamma=1,~M=1$. For simulation, we have taken $\epsilon = 0.001$.}
\label{ak-fig-1}  
\end{figure*}
\section{Derivation of $f_{\alpha}(z)$ for finite $z$}
\label{saddle-f}
In this appendix, we provide the derivation of the scaling function $f_{\alpha}(z)$ when $z$ is finite and which is written in Eq. \eqref{alph-neq-eq-14}. We rewrite $P(T_{loc},t)$ of Eq. \eqref{alph-neq-eq-11} in the form of Bromwich integral as
\begin{align}
P(T_{loc},t) \simeq \int _{E-i \infty}^{E+i \infty} \frac{ds}{2 \pi i} \frac{\mathcal{C}_{\alpha}}{s^{\frac{1}{2+\alpha}}} e^{t \Phi \left(s, \frac{\mathcal{C}_{\alpha} T_{loc}}{t}\right)},
\label{appen-f-alph-asy-eq-1}
\end{align}
where $\mathcal{C}_{\alpha}$ is given by Eq. \eqref{alph-neq-eq-911} and $\Phi(s,w)$ is defined as
\begin{align}
\Phi(s,w) = s - w s^{\frac{1+\alpha}{2+\alpha}},
\label{appen-f-alph-asy-eq-2}
\end{align}
where $w = \frac{\mathcal{C}_{\alpha} T_{loc}}{t}$.
At large $t$, the integral in Eq. \eqref{appen-f-alph-asy-eq-1} will be dominated by the saddle point of $\Phi(s,w) $ in $s$. The saddle point is given by $\left(\frac{d \phi}{ds}\right)_{s_*}=0$, solving which we find
\begin{align}
s_* = \left( \frac{1+\alpha}{2+\alpha}\right)^{2+\alpha} ~w^{2+\alpha}.
\label{appen-f-alph-asy-eq-3}
\end{align}
Note that $w$ is kept fixed throughout the derivation. We now expand $\Phi(s,w)$ about $s=s_*$ and substitute in Eq. \eqref{appen-f-alph-asy-eq-1} to get
\begin{align}
P(T_{loc},t)& \simeq \frac{e^{t \Phi(s_*, w)}}{2 \pi i} \frac{\mathcal{C}_{\alpha}}{s_*^{\frac{1}{2+\alpha}}} \int _{E-i\infty}^{E+i\infty} ds~ e^{t \frac{\Phi ''(s_*,w)}{2} (s-s_*)^2}, \nonumber\\
& \simeq\frac{e^{t \Phi(s_*, w)}}{2 \pi } \frac{\mathcal{C}_{\alpha}}{s_*^{\frac{1}{2+\alpha}}}  \int _{-\infty+i(s_*+E)}^{\infty+i(s_*+E)} dy~ e^{-t \frac{\Phi ''(s_*,w)}{2} y^2},\nonumber \\
& \simeq\frac{e^{t \Phi(s_*, w)}}{2 \pi } \frac{\mathcal{C}_{\alpha}}{s_*^{\frac{1}{2+\alpha}}}  \int _{-\infty}^{\infty} du~ e^{-t \frac{\Phi ''(s_*,w)}{2} u^2}.
\label{appen-f-alph-asy-eq-3}
\end{align}
We have used the notation $w = \frac{\mathcal{C}_{\alpha} T_{loc}}{t}$ and $\Phi''(s, w) = \frac{d^2 \Phi}{ds^2}$. While going from first line to second line, we have substituted $s-s_* = i y$ and while going to the third line, we have used that $\Phi''(s_*, w) = \frac{(2+\alpha)^{1+\alpha}}{(1+\alpha)^{2+\alpha}} \frac{1}{w^{2+\alpha}} $ is always greater than zero. This also implies that the integral is third line is always convergent. Performing this integral explicitly, we find the $P(T_{loc},t)$ obeys the scaling form
\begin{align}
P(t_{loc},t) \simeq \frac{\mathcal{C}_{\alpha}}{t^{\frac{1+\alpha}{2+\alpha}}}~f_{\alpha}\left(\frac{\mathcal{C}_{\alpha} T_{loc}}{t^{\frac{1+\alpha}{2+\alpha}}} \right)
\label{appen-f-alph-asy-eq-4}
\end{align}
where $f_{\alpha}(z)$ is given by
\begin{align}
f_{\alpha}(z) \simeq \frac{z^{\frac{\alpha}{2}}}{\sqrt{2 \pi}} \frac{(1+\alpha)^{\frac{\alpha}{2}}}{(2+\alpha)^{\frac{\alpha-1}{2}}} ~ \text{exp}\left( -\frac{(1+\alpha)^{1+\alpha}}{(2+\alpha)^{2+\alpha}}~ z^{2+\alpha}\right).
\label{appen-f-alph-asy-eq-5}
\end{align}
We emphasise that this expression is valid only for finite $z$ and breaks down for small values of $z$. The result in Eq. \eqref{appen-f-alph-asy-eq-5} is quoted in Eq. \eqref{alph-neq-eq-14} in the main text.

\section{Numerical verification of Eqs. \eqref{ak-eq-1} and \eqref{ak-eq-2}}
\label{appen-fig}
This appendix provides the numerical verification of Eqs. \eqref{ak-eq-1} and \eqref{ak-eq-2} which correspond to the distributions of $T_{loc}$ in presence of an absorbing wall at $x=M$ with initial position $x_0 = 0$ and $x_0 = -\epsilon$ respectively for $\alpha = 0$. The numerical verification is shown in Figure \ref{ak-fig-1}.


\end{document}